\documentclass[conference]{IEEEtran}
\IEEEoverridecommandlockouts
\usepackage{adjustbox}
\usepackage{cite}
\usepackage{amsmath,amssymb,amsfonts}
\usepackage[tight,footnotesize]{subfigure}
\usepackage{epsfig}
\usepackage{algorithm}
\usepackage{algorithmic}
\usepackage{graphicx}
\usepackage{textcomp}
\usepackage{xcolor}
\usepackage{comment}
\def\BibTeX{{\rm B\kern-.05em{\sc i\kern-.025em b}\kern-.08em
    T\kern-.1667em\lower.7ex\hbox{E}\kern-.125emX}}
    
\newcommand{\projectName}{UFZ}   
\usepackage{tikz}
\newcommand*\circled[1]{\tikz[baseline=(char.base)]{
            \node[shape=circle,fill,inner sep=0.5pt] (char) {\textcolor{white}{#1}};}}  
    
\begin{document}

\title{SZx: an Ultra-fast Error-bounded Lossy Compressor for Scientific Datasets\\
}

\author{\IEEEauthorblockN{Xiaodong Yu\IEEEauthorrefmark{1},
Sheng Di\IEEEauthorrefmark{1}, Kai Zhao\IEEEauthorrefmark{2},
jiannan Tian\IEEEauthorrefmark{3}, Dingwen Tao\IEEEauthorrefmark{3},
Xin Liang\IEEEauthorrefmark{4}, Franck Cappello\IEEEauthorrefmark{1}}

\IEEEauthorblockA{\IEEEauthorrefmark{1}Argonne National Laboratory, Lemont, IL\\
\IEEEauthorrefmark{2}University of California, Riverside, CA\\
\IEEEauthorrefmark{3}Washington State University, Pullman, WA\\
\IEEEauthorrefmark{4}Missouri University of Science and Technology, Rolla, MO\\
\{xyu, sdi1\}@anl.gov, kzhao016@ucr.edu, \{Jiannan.tian, dingwen.tao\}@wsu.edu, xliang@mst.edu, cappello@mcs.anl.gov}
\thanks{Corresponding author: Sheng Di, Mathematics and Computer Science Division, Argonne National Laboratory, 9700 Cass Avenue, Lemont, IL 60439, USA}
}

\maketitle

\begin{abstract}
Today's scientific high performance computing (HPC) applications or advanced instruments are producing vast volumes of data across a wide range of domains, which introduces a serious burden on data transfer and storage. Error-bounded lossy compression has been developed and widely used in scientific community, because not only can it significantly reduce the data volumes but it can also strictly control the data distortion based on the use-specified error bound.  
Existing lossy compressors, however, cannot offer ultra-fast compression speed, which is highly demanded by quite a few applications or use-cases (such as in-memory compression and online instrument data compression). In this paper, we propose a novel ultra-fast error-bounded lossy compressor, which can obtain fairly high compression performance on both CPU and GPU, also with reasonably high compression ratios. The key contributions are three-fold: (1) We propose a novel, generic ultra-fast error-bounded lossy compression framework -- called {\projectName}, by confining our design to be composed of only super-lightweight operations such as bitwise and addition/subtraction operation, still keeping a certain high compression ratio. (2) We implement {\projectName} on both CPU and GPU and optimize the performance according to their architectures carefully. (3) We perform a comprehensive evaluation with 6 real-world production-level scientific datasets on both CPU and GPU. Experiments show that {\projectName} is 2$\sim$16$\times$ as fast as the second-fastest existing error-bounded lossy compressor (either SZ or ZFP) on CPU and GPU, with respect to both compression and decompression.  
\end{abstract}


\section{Introduction}
With ever-increasing complexity of modern scientific research, today's high performance computing (HPC) applications and advanced instruments are producing extremely large volumes of data in their simulations or experiments. Hardware/Hybrid Accelerated Cosmology Code (HACC) \cite{hacc}, for example, could produce 20TB of simulation data at only one run with hundreds of simulation iterations and trillions of particles to involve.  

During the past five years, quite a few outstanding error-bounded lossy compressors have been developed to resolve the big data issue, however, the compression/decompression throughput is still far lower than the target performance demanded by many use-cases, such as instrument data compression and in-memory compression. Linear Coherent Light Source (LCLS-II) \cite{lcls-ii} could generate the instrument data at a rate of 250GB/s \cite{use-case}, and these data need to be stored and transferred to a parallel file system (FPS) timely for post hoc analysis. By comparison, the single-core CPU performance of existing lossy compressors is generally only 200$\sim$400MBs \cite{sz17,Xin-bigdata18} and the GPU performance is only 10$\sim$66GB/s \cite{cusz,cuZFP}, which has also been verified in our experiments. Another typical example is exa-scale parallel quantum computing (QC) simulation, which requires a fairly large memory capacity (e.g., $2^{58}$$\approx$256PB when simulating 50 qubits each with a double precision) for each run in practice \cite{qc-comp}. In order to reduce memory requirement significantly, the QC simulation researchers \cite{qc-comp} have developed a method to store the lossy-compressed data in memory and decompress them whenever needed in the course of the simulation, while suffering from considerable overhead in simulation time (even up to $\sim$20$\times$ in worst case), which is totally undesired by users.

In this paper, we focus on how to significantly accelerate both compression and decompression performance for error-bounded lossy compression while keeping a certain high compression ratio, which faces two grand challenges to resolve. On the one hand, in order to pursue the ultra-high lossy compression/decompression performance, we have to strict the whole design to be limited to only super-lightweight operations including addition/subtraction/bitwise operations, which is a serious challenge. Specifically, the relatively expensive operations such as multiplication and division should be suppressed because of its significantly higher cost. All of the existing efficient error-bounded lossy compressors, however, depend on such expensive operations. For instance, SZ2.1 \cite{Xin-bigdata18} relies on the linear regression prediction, which involves masses of multiplications to compute the coefficients. Moreover, SZ2.1 relies on a linear-scale quantization to control the user-specified error bound, which involves a division operation (\textit{quantzation\_bin}=$[\frac{prediction\_error}{2\cdot error\_bound}+\frac{1}{2}]$ \cite{tpds-point-wise}) on each data point. ZFP \cite{zfp} is another state-of-the-art error-bounded lossy compressor, which is designed based on the data transform, also involving masses of matrix-multiplication operations. On the other hand, how to optimize the performance to adapt to different device architectures is very challenging, because it requires fairly detailed and non-trivial performance tuning in the regard of practical experiments with numerous real-world scientific datasets across different domains. 

To address the above serious challenges, we propose a novel, \textbf{U}ltra-\textbf{F}ast error-bounded lossy \textbf{C}ompression framework -- namely \textit{\projectName}, which can also be extended/implemented efficiently for different devices such as CPU and GPU. The key contributions are summarized as follows: 
\begin{itemize}
    \item We develop {\projectName}, which composes only lightweight operations such as bitwise operation, addition and subtraction. {\projectName} also supports strict control of the compression errors within user-specified error bound, thanks to our elaborate design in its error-control mechanism. 
    \item We implement {\projectName} for CPU and GPU accelerator, and also optimize their performances carefully based on these device architectures, respectively. 
    \item We comprehensively evaluate {\projectName} by running it with 6 real-world scientific datasets on heterogeneous compute nodes offered by different supercomputers including ORNL Summit and ANL ThetaGPU. We rigorously compare {\projectName} to two state-of-the-art lossy compressors  SZ and ZFP, as well as their GPU versions -- cuSZ, cuZFP. 
    \item Experiments show that {\projectName} is 2$\sim$5$\times$ as fast as the second-best existing error-bounded lossy compressor on CPU and 5$\sim$10$\times$ as fast as the second-best on GPU, with respect to both compression and decompression. At such a high performance, {\projectName} can still get a very nice compression ratio (3$\sim$12 for the overall compression ratio of each application; and up to 124 for the compression ratio of specific field) with good reconstructed data quality. 
\end{itemize}

The rest of the paper is organized as follows. In Section \ref{sec:relate}, we discuss related work. In Section \ref{sec:design}, we present the design overview of our ultra-fast error-bounded lossy compression framework. In Section \ref{sec:optimization}, we propose optimization strategies in improving the performance for different devises including CPU and GPU. In Section \ref{sec:evaluation}, we present and discuss the performance evaluation results. Finally, we conclude the paper with a vision of the future work in Section \ref{sec:con}.

\section{Related Work}
\label{sec:relate}

%Compression/decompression performance is one of the most important metrics in the compression assessment. Liang et al. \cite{Xin-Cluster-data-dump} presented that the compression/decompression speed could be a dominant factor for the overall compressed data dumping performance especially when the system has a relatively high I/O bandwidth. Wu et al \cite{qc-comp} also showed that large-scale quantum computing simulation requires not only high compression ratios but also very high compression and decompression speed in particular, in order to prevent the simulation from being executed with a prohibitively long time.   

Basically, high-speed scientific data compression can be split into two categories - lossless compression and lossy compression, which will be discussed in the following text, especially in the regard of performance/speed. 

High-speed lossless compressors have been developed in particular because of the strong demand on compression performance in many use-cases. Facebook Zstd \cite{zstd}, for example, was developed particularly for the sake of high performance, with very similar compression ratios compared with other state-of-the-art lossless compressors such as Zlib \cite{zlib} and Gzip \cite{gzip}. In general, Zstd could be 5$\sim$6$\times$ faster than Zlib as shown in \cite{zstd}, such that it has been widely integrated/used in 80+ production-level software/libraries/platforms. 
Unfortunately, Zstd supports only lossless compression, which would suffer very low compression ratios (1.2$\sim$2 in most of cases) when compressing scientific datasets that are mainly composed of floating-point values (to be shown later). 

High-speed lossy compression has also gained significant attentions by compressor developers or scientific applications/researchers. SZ \cite{sz16,sz17,Xin-bigdata18} is a typical fast error-bounded lossy compressor, which can reach 200$\sim$300MB/s in compression and decompression speed \cite{sz16,sz17, Xin-bigdata18}. However, it is still not as fast as expected by the quantum computing simulations \cite{qc-comp}, so a faster lossy compression method (namely \textit{QCZ} in the following text) was customized with comparable compression ratios (especially for a high-precision compression with relative error bound of 1E-4 or 1E-5). ZFP \cite{zfp} is another fast error-bounded lossy compressor, which is well known for its relatively high compression ratios and fairly high compression speed in both CPU and GPU. Based on our experiments (to be shown later), ZFP and QCZ has comparable compression speed, and they are generally 1.5$\sim$2$\times$ as fast as SZ. In fact, SZ already has a fairly high performance compared with other compressors as demonstrated in literature: it has a comparable performance with FPZIP \cite{fpzip} and SZauto \cite{szauto} and about one to two orders of magnitude higher performance than ISABELA \cite{isabela}, MGARD \cite{mgard} and TTHRESH \cite{tthresh}. 

Because of the high demand on ultra-fast error-bounded lossy compressors, a few specific error-controlled lossy compression algorithms have been developed for GPU accelerators in particular, and cuSZ \cite{cusz} and cuZFP \cite{cuZFP} are two leading ones.
The cuSZ is the only GPU-based lossy compressor supporting absolute error bound for scientific data compression. It was designed based on the classic prediction-based compression model SZ and optimized for GPU performance significantly by leveraging a dual-quantization strategy \cite{cusz} to deal with the Lorenzo prediction. Since ZFP's core stage is performing a customized orthogonal data transform that can be executed in the form of matrix-multiplication, cuZFP can leverage high-performance CUDA library to reach a very high throughput. CuZFP, however, does not support error-bounded compression but only fixed-rate compression, which suffers from very low compression ratios, as verified in \cite{fraz}.  

In comparison with all the above related works, our proposed {\projectName} is about 2$\sim$5$\times$ as fast as the second-fastest lossy compressor ZFP on CPU and 2$\sim$10$\times$ as fast as the second-fastest (cuSZ) on GPU, also with relatively high compression ratios (3$\sim$12 depending on user's error bound).

\section{Problem Formulation}
\label{sec:problem}

In this section, we formulate the research problem we focus on in this paper: optimization of the error-bounded lossy compression/decompression performance with as high compression ratios as possible. Specifically, given a scientific dataset (denoted by $D$) composed of $N$ data values each denoted by $d_i$, where $i$=1,2,3,$\cdots$,$N$. The objective of our work is to develop an error-bounded lossy compressor with an ultra high performance in both compression and decompression for both CPU and GPU, also strictly respecting the user-required error bound, which can be represented as the following formula.

\begin{equation}
\label{eq:formulation}
\begin{array}{l}
 \max (CT)\hspace{1mm}and\hspace{1mm}\max (DT) \\ 
 s.t.\hspace{1mm}|d_i - d_i'| \le e \\ 
 \hspace{6mm}C\hspace{-0.2mm}R\hspace{0.7mm}is\hspace{0.7mm}relatively\hspace{0.7mm}high\\ 
 \end{array}
\end{equation}
where \emph{CT} and \emph{DT} represent the compression throughput and decompression throughput, respectively; $d_i$ and $d_i'$ denote the original data value and decompressed data value in the dataset, respectively; $e$ is referred to as the user-specified absolute error bound, and \emph{CR} means the compression ratio, which is defined as the ratio of the original data size to the lossy compressed data size. In order to obtain as high performance as possible, the compression ratio (CR) would definitely be not optimal. However, we still hope to get a relatively high CR (expected to be over 5 or 10), which is still much higher than lossless compression ratio (generally 1.2$\sim$2 for scientific data).

The compression throughput and decompression throughput are defined in Formula (\ref{eq:ct}) and Formula (\ref{eq:dt}), respectively.
\begin{equation}
\label{eq:ct}
C\hspace{-0.1mm}T = (N \cdot b)/{T}
\end{equation}
\vspace{-5mm}
\begin{equation}
\label{eq:dt}
D\hspace{-0.1mm}T = (N \cdot b)/{T'}
\end{equation}
where $N$ is the number of data points in the dataset $D$, $b$ represents the number of bytes per value in $D$ (e.g., $b$ = 4 when the original data precision is single-precision floating point); $T$ and $T'$ denote the time cost when compressing the dataset $D$ and the time cost when decompressing the corresponding compressed data, respectively.

In addition to the maximum compression error (i.e., error bound as shown in Formula (\ref{eq:formulation})), we will also evaluate the reconstructed data quality by commonly used statistical data distortion metrics such as Peak Signal to Noise Ratio (PSNR) \cite{z-checker} and Structural Similarity Index Measure (SSIM) \cite{ssim}, which have been commonly used in lossy compression and visualization community. In general, the higher the PSNR or the higher the SSIM, the better the reconstructed data quality.

\section{Ultra-fast Error-bounded Lossy Compression Framework -- {\projectName}}
\label{sec:design}

In this section, we present the design overview of our ultra-fast error-bounded lossy compression framework -- {\projectName}. Detailed performance optimization strategies for CPU and GPU will be discussed in next section.

\begin{figure}[ht] \centering

\hspace{-8mm}
\subfigure[{Miranda (pressure:slice128)}]
{
\raisebox{-1cm}{\includegraphics[scale=0.25]{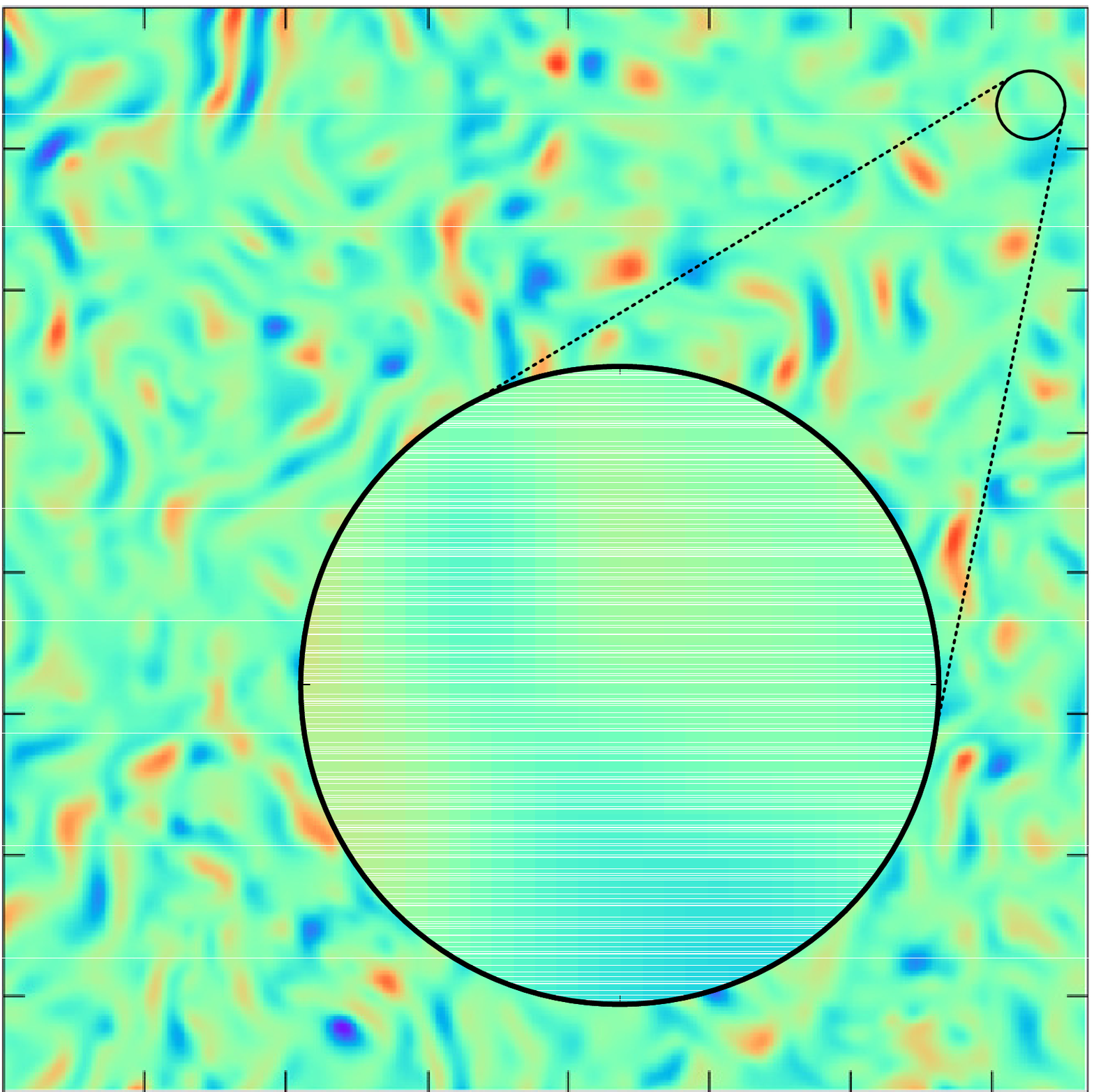}}
}
\hspace{-3mm}
\subfigure[{Nyx cosmology (temperature)}]
{
\raisebox{-1cm}{\includegraphics[scale=0.25]{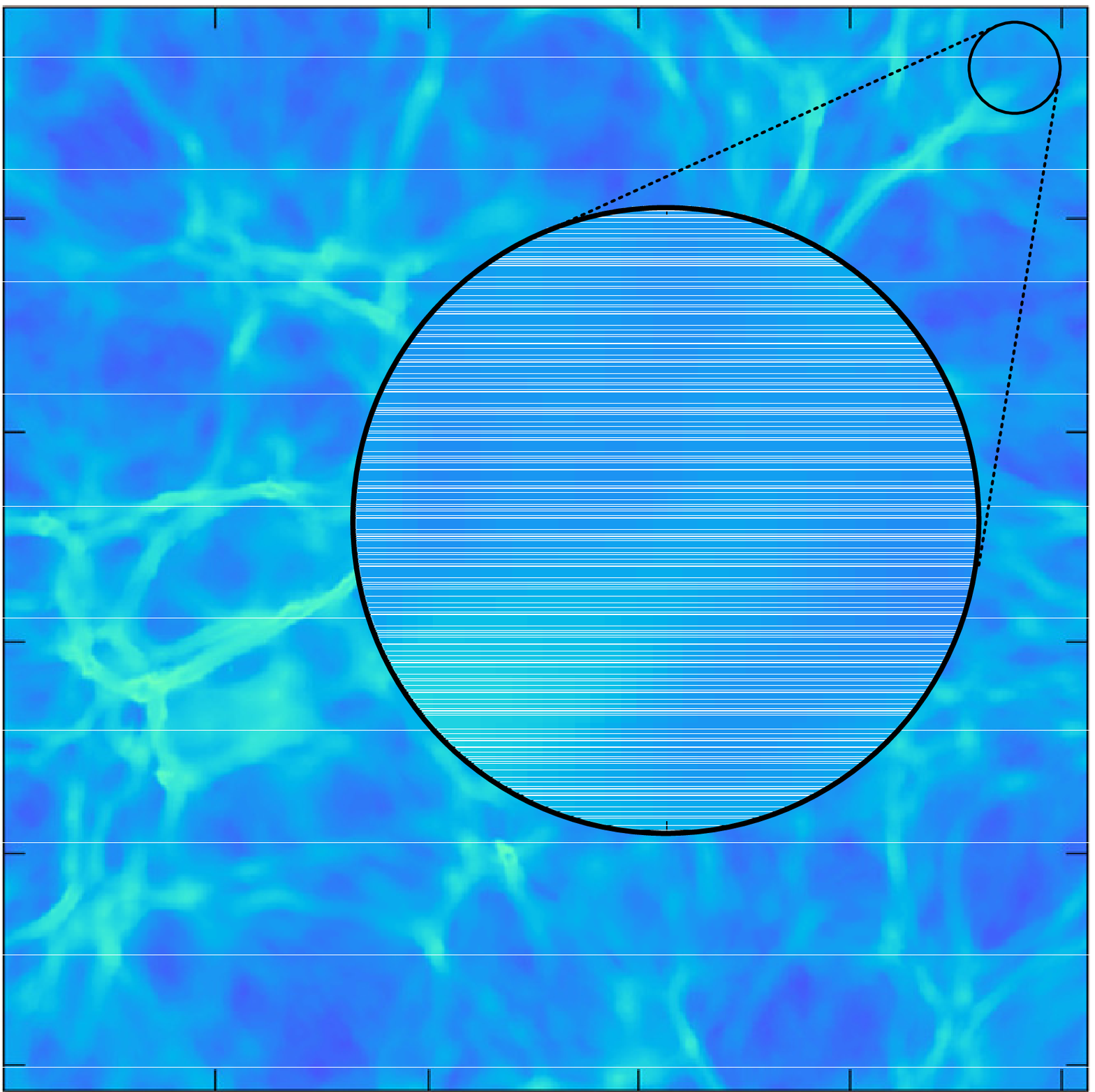}}
}
\hspace{-10mm}

\hspace{-7mm}
\subfigure[{QMCPack (slice500)}]
{
\raisebox{-1cm}{\includegraphics[scale=0.25]{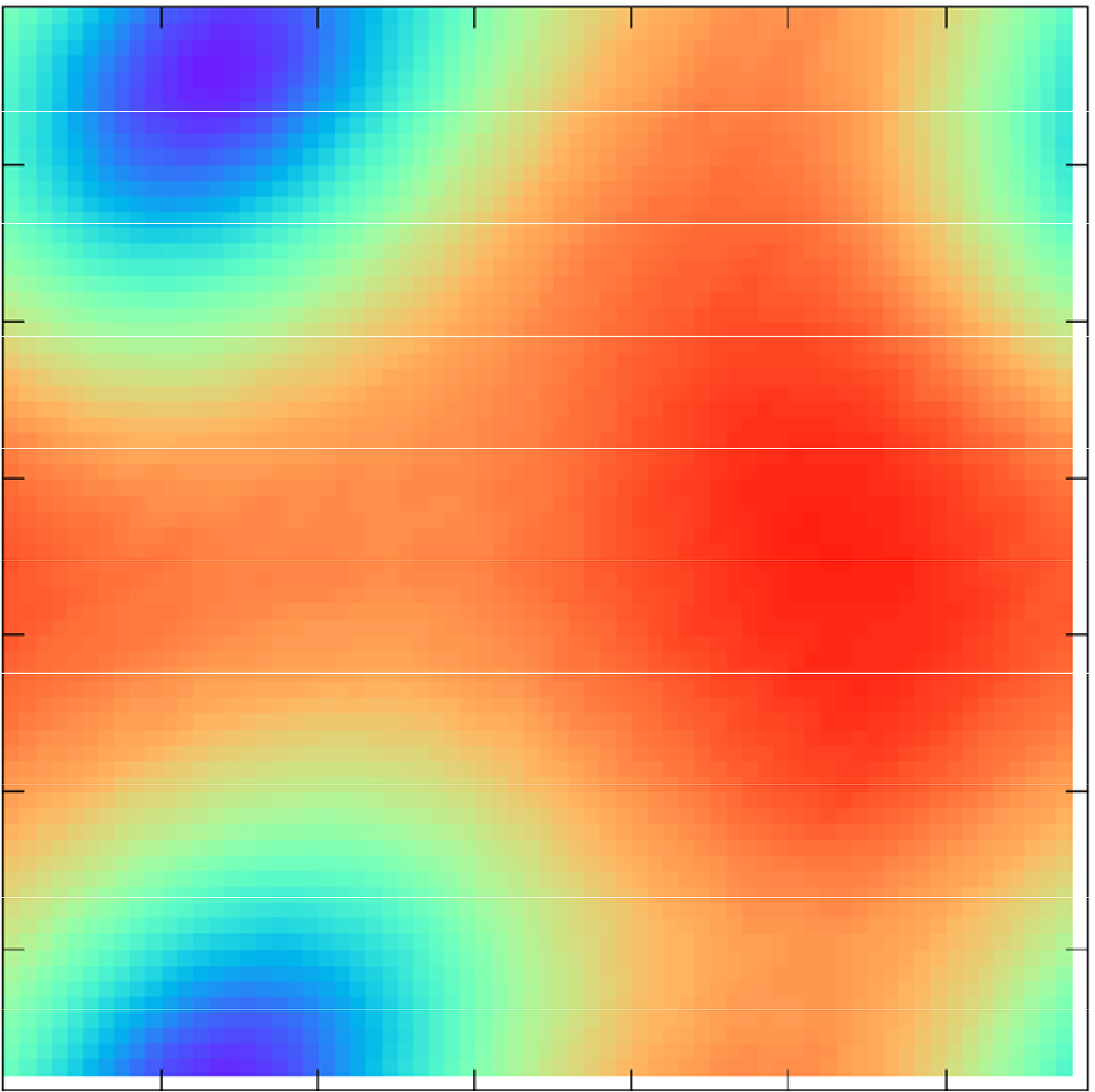}}
}
\hspace{-3mm}
\subfigure[{Hurricane (U:slice60)}]
{
\raisebox{-1cm}{\includegraphics[scale=0.25]{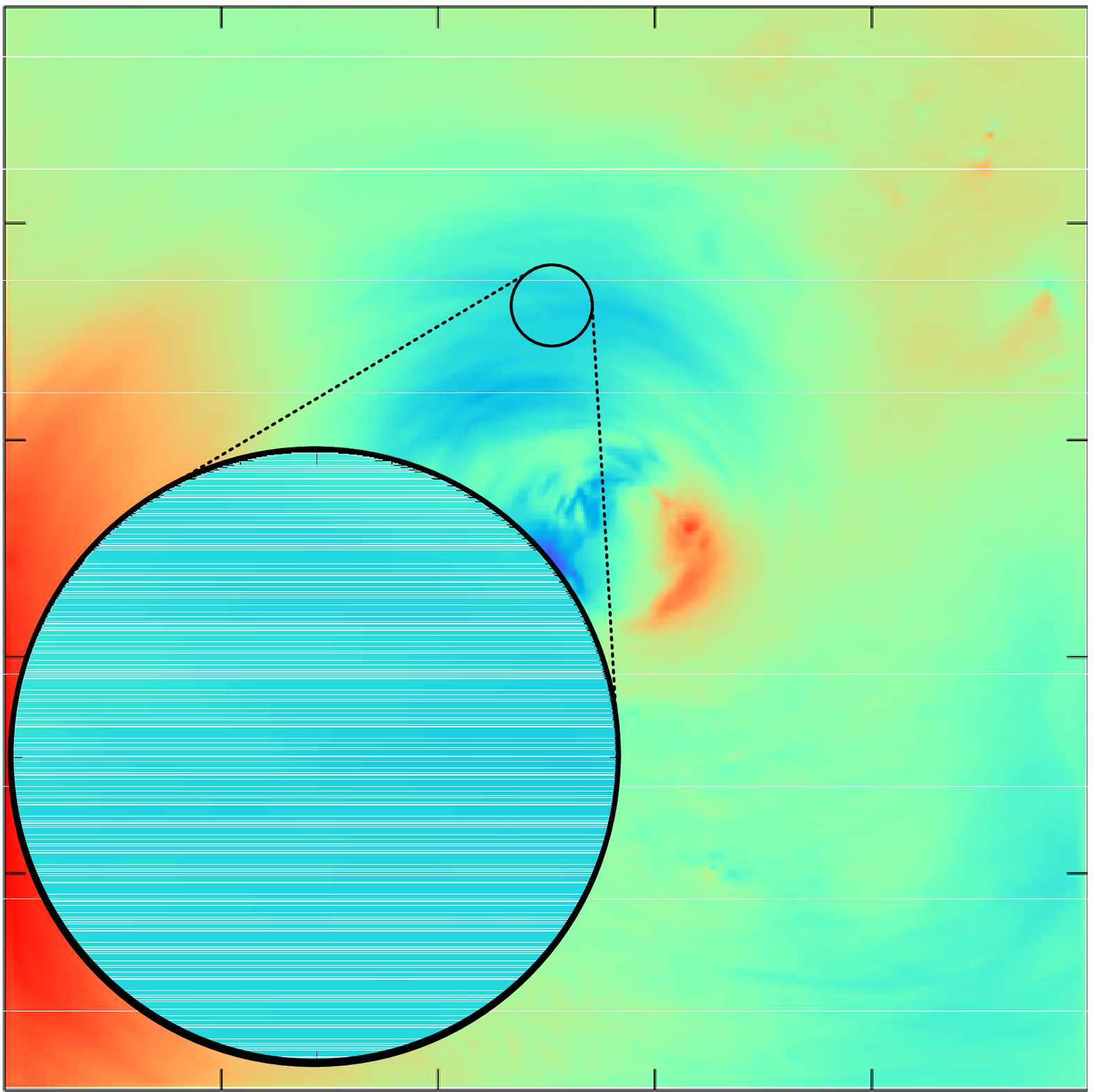}}
}
\hspace{-10mm}
\vspace{-1mm}

\caption{Demonstrating High Smoothness of Scientific Datasets}
\label{fig:vis-smoothness}
\end{figure}

\begin{figure}[ht] \centering

\hspace{-8mm}
\subfigure[{Miranda (pressure:slice128)}]
{
\raisebox{-1cm}{\includegraphics[scale=0.39]{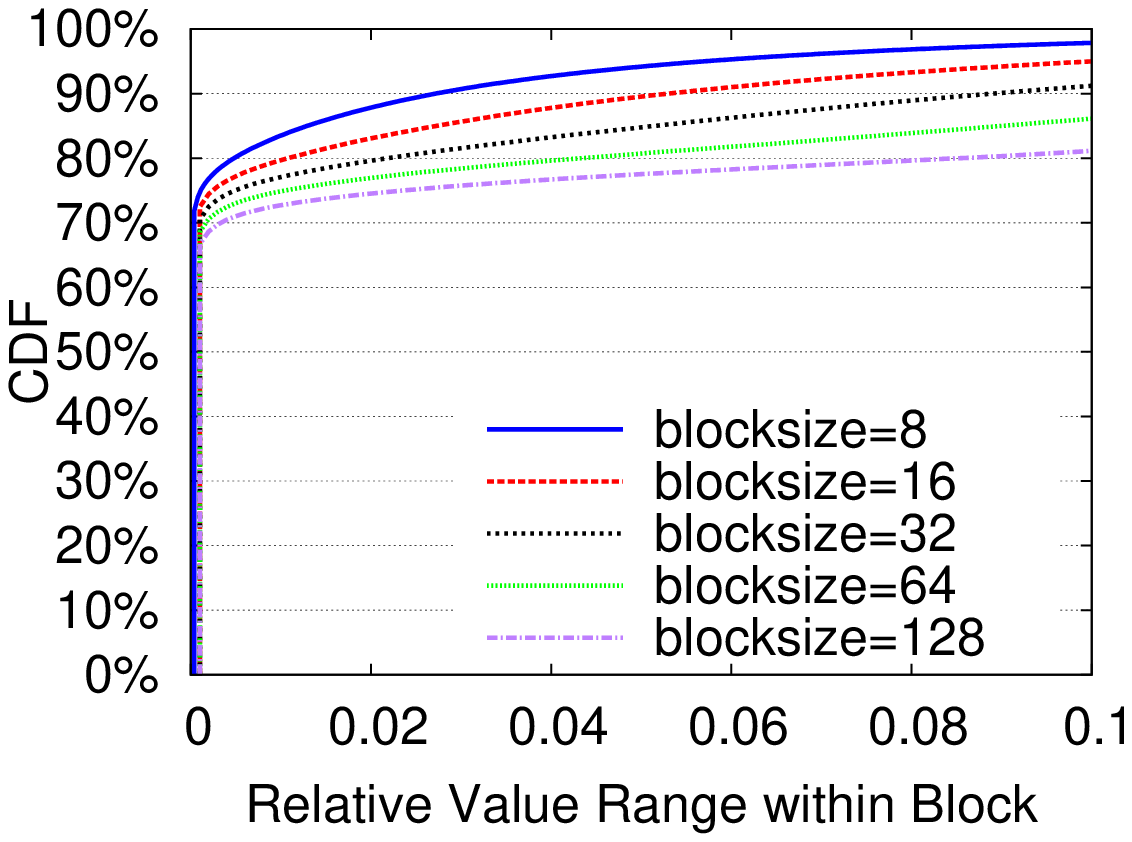}}
}
\hspace{-10mm}
\subfigure[{Nyx cosmology (temperature)}]
{
\raisebox{-1cm}{\includegraphics[scale=0.39]{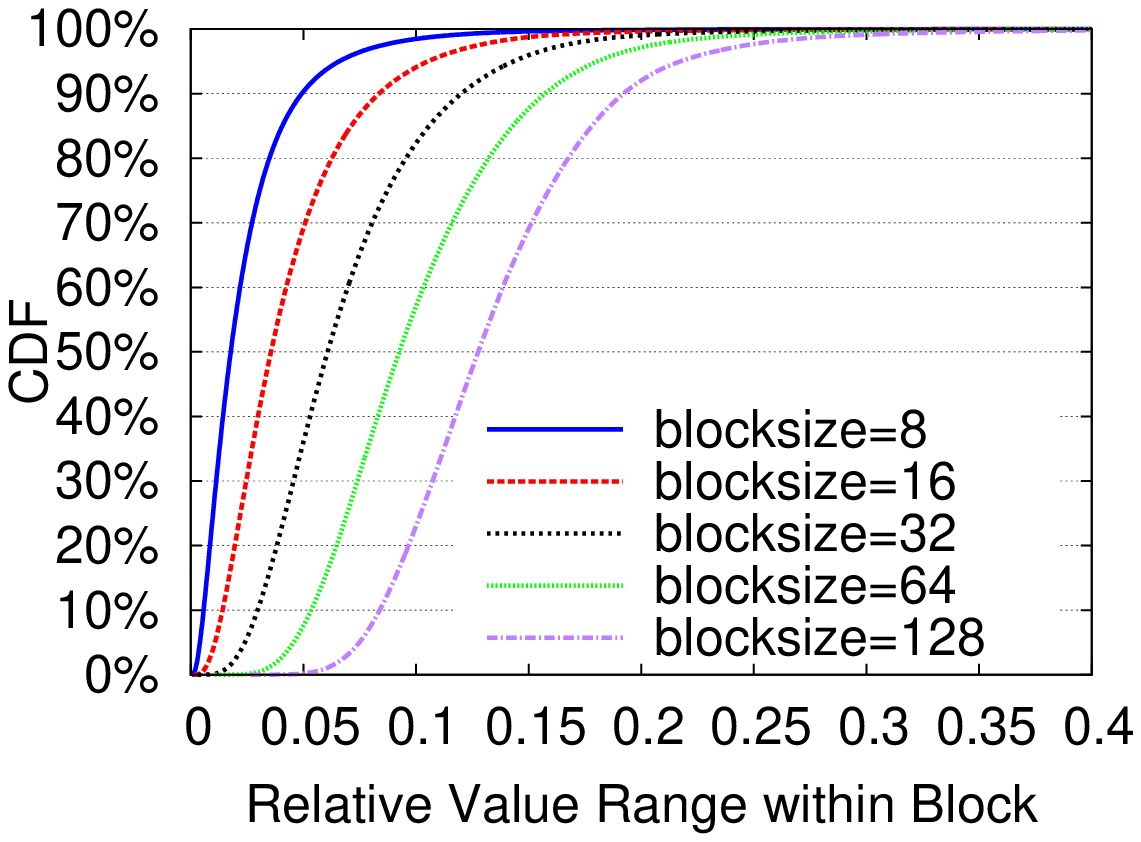}}
}
\hspace{-10mm}

\hspace{-8mm}
\subfigure[{QMCPack (slice500)}]
{
\raisebox{-1cm}{\includegraphics[scale=0.39]{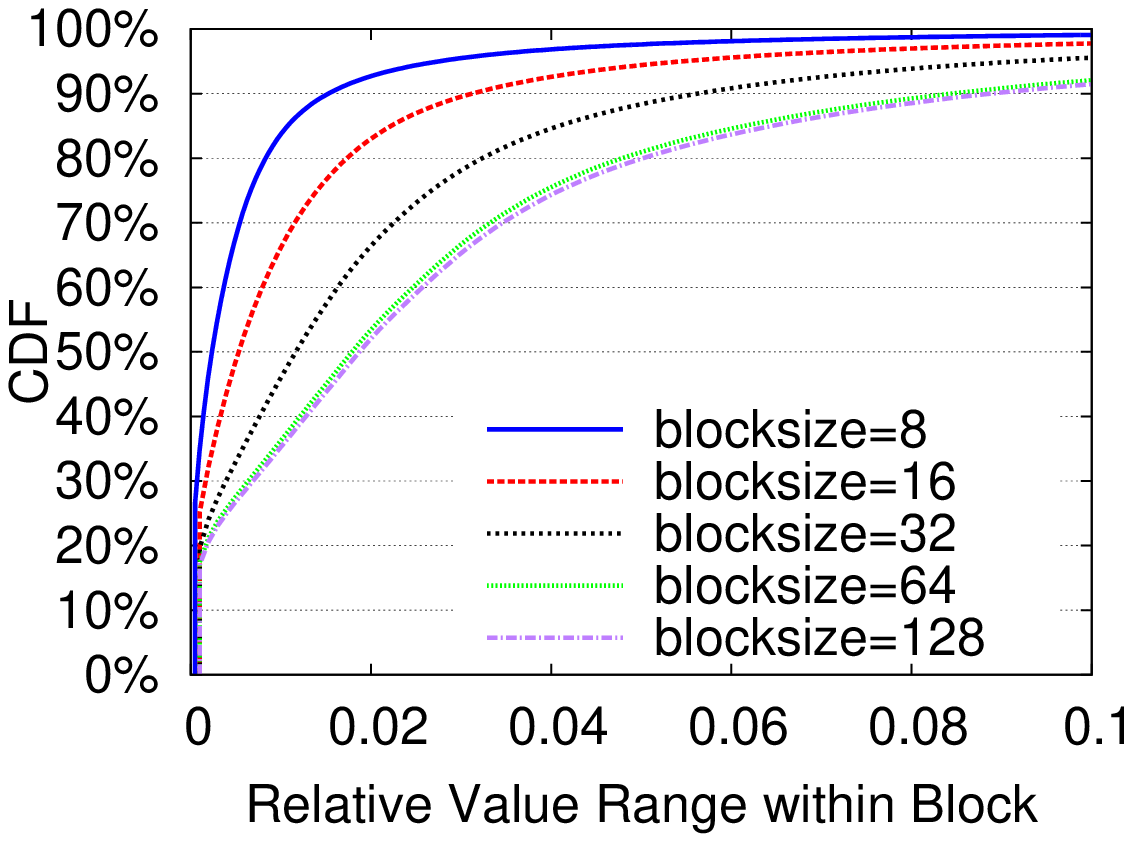}}
}
\hspace{-10mm}
\subfigure[{Hurricane (U:slice60)}]
{
\raisebox{-1cm}{\includegraphics[scale=0.39]{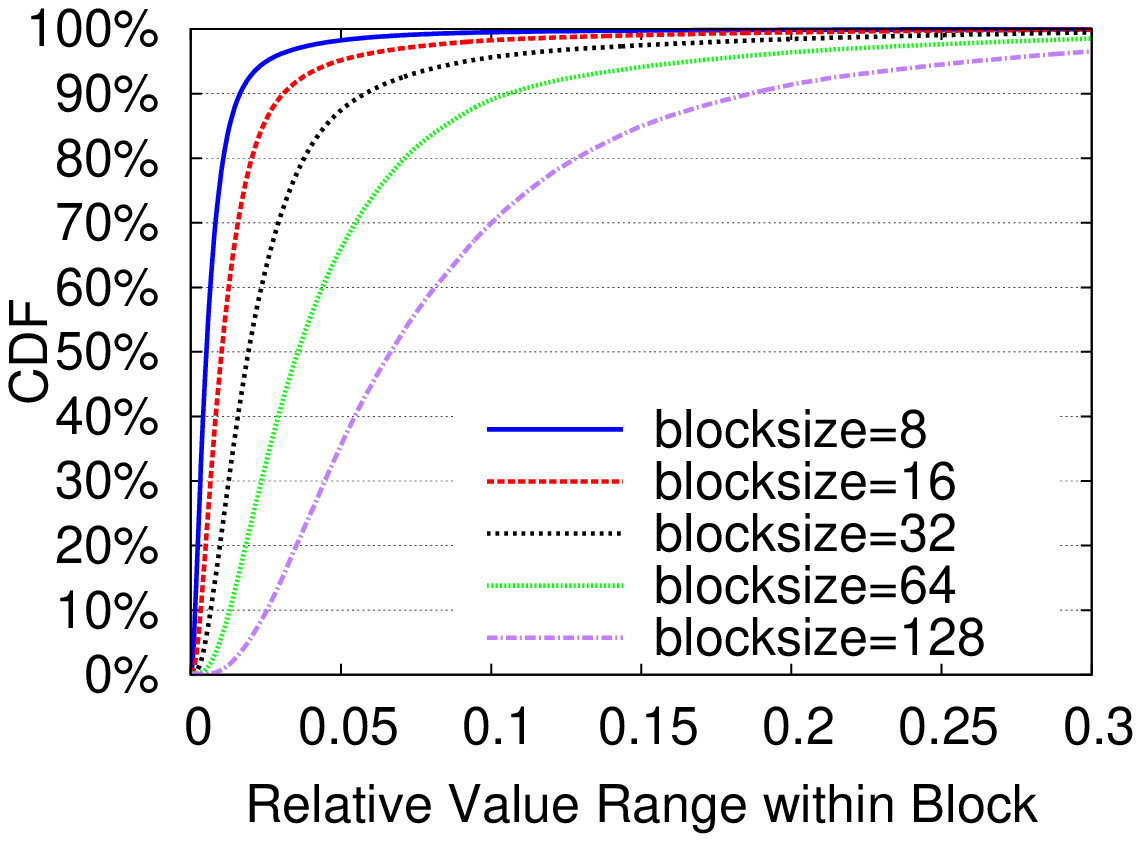}}
}
\hspace{-10mm}
\vspace{-1mm}

\caption{Cumulative Distribution Function (CDF) of Block's Value Range}
\label{fig:cdf}
\end{figure}

Our design is motivated by the fact that most of the scientific datasets are pretty smooth in space, such that all the values in a small block (e.g., 16 or 32 consecutive data points) are likely very close to each other, thus the mean of the minimal value and maximal value in the block can be used to represent the whole block based on a certain error bound. Figure \ref{fig:vis-smoothness} shows the visualization of four typical fields across from four different real-world simulation datasets (Miranda large-eddy simulation \cite{Miranda}, Nyx cosmology simulation \cite{nyx}, QMCPack quantum chemistry \cite{qmcpack}, and Hurricane climate simulation \cite{hurricane-2004}), clearly demonstrating the high smoothness of the data in local spatial regions. Furthermore, Figure  \ref{fig:cdf} shows the cumulative distribution function (CDF) of block's relative value range\footnote{A block's relative value range is defined as the ratio of the block's value range to the dataset's global value range. The reason we check the block's relative value range is that the error-bounded lossy compression is often performed via value-range based relative error bound \cite{z-checker}, where the absolute error bound is calculated based on the dataset's global value range.}. It verifies that the four scientific datasets all exhibit fairly high smoothness of the local data without loss of generality. Specifically, for the Miranda dataset and QMCPack dataset, 80+\% of blocks have very small relative value ranges ($\leq$0.01), when the block size is 8. 

\begin{figure}[ht]
    \centering
    \includegraphics[width=1\columnwidth]{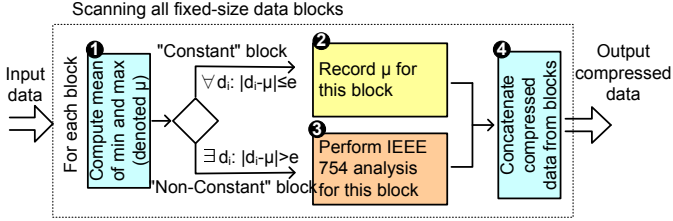}
    \caption{Design Architecture/Workflow of {\projectName}}
    \label{fig:workflow}
\end{figure}

We design our compressor {\projectName} in terms of the local smoothing feature, as illustrated in Figure \ref{fig:workflow}. The fundamental idea is organizing the whole dataset as many small 1D blocks (or segments) and checking whether the mean of min and max (denoted by $\mu$) in each block can be used to represent all values in this block with deviations respecting user-specified error bound. If yes, we call this block `constant' block, so we just need to store $\mu$ for this block of data; or else, we  compress all the data points in this block by analyzing their IEEE 754 representations in terms of the user-required error bound. 

We present the pseudo-code of the skeleton design in Algorithm \ref{alg:design-skeleton} to further describe details. Table \ref{tab:notation} summarizes all key notations to assist understanding of the algorithm. 
\begin{table}[h]
\centering
\caption{Key Notations Involved in The ULF Algorithm}
\begin{tabular}{|c|l|}
\hline
\textbf{Notation} & \textbf{Description} \\ \hline
$D$ & The dataset given for compression \\ \hline
$e$ & user-specified error bound
\\ \hline
$d_i$ & the data points in the original raw dataset $D$
\\ \hline
$B_k$ & $k$th block in the dataset
\\ \hline
$\mu_k$ & mean of min and max in Block $k$
\\ \hline
$r_k$ & variation radius of Block $k$ 
\\ \hline
$R_k$ & the required mantissa bits calculated via $e$ and $\mu_k$ for $B_k$
\\ \hline
$v_i$ & normalized values based on $\mu_k$ in each block $B_k$
\\ \hline
$L_i$ & identical leading bits of $v_i$ compared with $v_{i-1}$
\\ \hline
\end{tabular}
\label{tab:notation}
\end{table}

We describe Algorithm \ref{alg:design-skeleton} as follows.
As mentioned previously, the whole dataset is split into many small fixed-size 1D blocks and the compression will be executed block by block (line 2). Because of the high smoothness of data in locality, quite a few data blocks may have the values already respect the error bound based on the mean of min and max (denoted by $\mu$) (line 4$\sim$6), and such blocks are called `constant' blocks, which will be compressed by simply storing the corresponding $\mu$ value. 
\begin{algorithm}
\caption{\textsc{Skeleton Design of {\projectName}}} \label{alg:design-skeleton} \footnotesize
\renewcommand{\algorithmiccomment}[1]{/*#1*/}
\begin{flushleft}
\textbf{Input}: dataset $D$, user-specified error bound $e$, block size (denoted $b$)\\
\textbf{Output}: compressed data stream in form of bytes
\end{flushleft}

\begin{algorithmic} [1]
\STATE $i$ $\leftarrow$ 0, $k$ $\leftarrow$ 0;\COMMENT{Set 0 to all counters} 
\FOR{each block $B_k$ with block size $b$}
    \STATE Compute $\mu_k$ for $B_k$; \COMMENT{Compute mean of min and max}
    \IF{($\forall$$d_i$$\in$$B_k$: $|d_i-\mu_k|\leq e$)}
        \STATE $\mu\_$\textit{array} $\leftarrow$ $\mu_k$; \COMMENT{Collect $\mu$ for `constant' blocks}
    \ELSE
        \STATE Compute required \# mantissa bits (denoted as $R_k$);
        \FOR{each normalized value $v_i$ in $B_k$}
            \STATE Compute \textit{identical\_leading\_bytes} for $v_i$ and $v_{i-1}$;
            \STATE Encode \textit{identical\_leading\_bytes} into \textit{xor\_leadingzero\_array};
            \STATE \textit{mb\_array} $\leftarrow$ $R_k$ $-$ $L_i$; \COMMENT{Commit required bits excluding $L$}
        \ENDFOR 
    \ENDIF
    \STATE Aggregate output: $\mu\_$\textit{array}, \textit{xor\_leadingzero\_array}, \textit{mb\_array};
\ENDFOR
\end{algorithmic}
\end{algorithm}

\begin{figure}
    \centering
    \includegraphics[width=1\columnwidth]{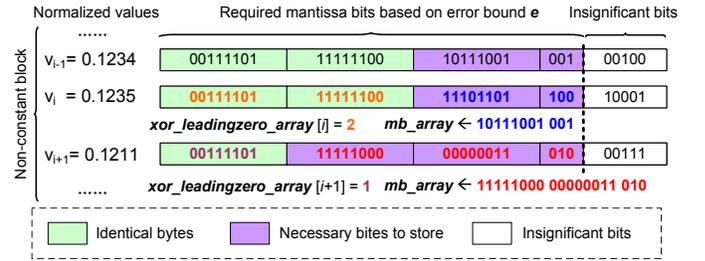}
    \caption{Compressing non-constant block by binary representation analysis: suppose three adjacent normalized values in a non-constant block are 0.1234, 0.1235, and 0.1211, respectively.}
    \label{fig:mb_array}
\end{figure}

For each of the non-constant blocks, we first normalize the data by subtracting the variation radius of the block (i.e, mean of min and max for the data in the block), and then compress each such normalized value by IEEE 754 binary representation analysis according to the following three steps: 
\begin{itemize}
    \item \textbf{Line 7}: We compute the required number of mantissa bits (denoted as $R_k$) based on user-specified error bound, by the following formula. 
\begin{equation}
\label{eq:required-bits}
\hspace{-12mm}
R_k \hspace{-1mm} = \hspace{-1mm}\left\{ \begin{array}{l}
 \hspace{-2mm}0,\hspace{19mm}p(r _k ) - p(e) \leq 0 \\ 
 \hspace{-2mm}sizeof(type),\hspace{1.5mm}p(r _k ) - p(e) > sizeof(type) \\ 
 \hspace{-2mm}p(r _k ) - p(e),\hspace{4mm}otherwise \\ 
 \end{array} \right.\hspace{-6mm}
\end{equation}
    where $p(x)$ denotes getting the exponent of the number $x$ , $r_k$ denotes the variation radius of data in the block $k$, and sizeof(type) refers to the length of the data type (e.g., 32bits for single-precision floating-point type).
    The idea is normalizing the data values by subtracting the mean of min and max, such that the maximum exponent of each normalized value is foreseeable and thus the required mantissa bits are estimable by combining the exponent of the error bound value $e$.
    \item \textbf{Line 9}: Compute identical leading bytes by an XOR operation between the normalized data value $v_i$ and its preceding data value $v_{i-1}$. The number of leading zeros after the XOR operation indicates the number of identical leading bytes between the two data points. 
    \item \textbf{Line 10}: We encode the number of identical leading bytes for each data point by a 2-bit code: 00, 01, 10, and 11 corresponds to 0, 1, 2, and 3 identical leading bytes, respectively. We use a 2-bit-per-value array (called \textit{xor\_leadingzero\_array}) to carry these 2-bit codes, as illustrated in Figure \ref{fig:mb_array}. 
    \item \textbf{Line 11}: We commit the necessary mantissa bits, i.e., error-bounded based required bits (denoted as $R_k$) excluding identical leading bytes (denoted by $L_i$, a.k.a, \textit{mid-bytes}), to a particular mantissa bit array (denoted as \textit{mb\_array}), as shown in Figure \ref{fig:mb_array}. 
\end{itemize}

\section{Performance Optimization for Various Devices}
\label{sec:optimization}

In this section, we describe our specific performance optimization strategies for CPU and GPU, respectively. 

\subsection{Performance Optimization for CPU}
\label{sec:perfopt}

In this subsection, we describe how to accelerate the {\projectName} in our CPU code by an efficient bitwise right shifting operation, which mainly involves the line 7$\sim$12 in Algorithm \ref{alg:design-skeleton}. This is a fundamental optimization strategy which can also be applied in other devices/accelerators such as GPU. In what follows, we first describe a potential performance issue in the {\projectName} design, followed by our optimization solution thereafter. 

\begin{figure}[ht]
    \centering
    \includegraphics[width=1\columnwidth]{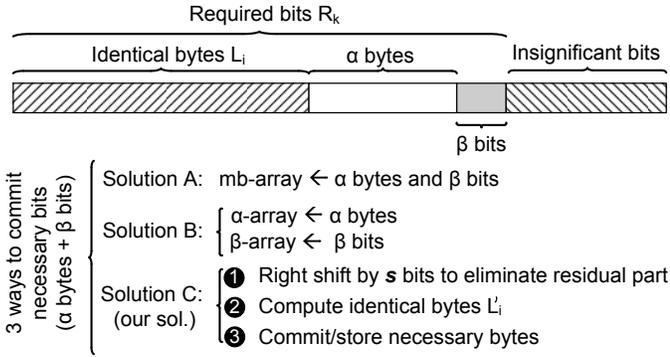}
    \caption{Illustration of 3 Ways to Store Necessary Mantissa Bits (Solution C is our performance optimization strategy)}
    \label{fig:cpu-perf-opt}
\end{figure}

As illustrated in Figure \ref{fig:cpu-perf-opt}, the mantissa bits that need to be stored for the normalized value $v_i$ should exclude the identical bytes $L_i$ and the insignificant bits which are calculated based on the user-specified error bound and variation radius of the corresponding block. The number of such necessary mantissa bits is generally not a multiple of 8 (to be verified later), so that committing/storing these bits in the compressed data requires specific bitwise operation strategies.   

Storing a short bit-array with an arbitrary number of bits is a very common operation in lossy compression. The most straight-forward solution (Solution A as shown in Figure \ref{fig:cpu-perf-opt}) is treating the given bit-array as a particular integer, and populate the target bit-stream pool (i.e., mb-array in the figure) by applying a couple of bit-wise operations (such as bit-shift, bit-and and bit-or) on the integer number. Many lossy compressors store the arbitrary bits in this way, such as Pastri \cite{pastri}. An alternative solution (Solution B as shown in Figure \ref{fig:cpu-perf-opt}) is splitting the necessary bits into two parts -- a number of necessary bytes ($\alpha$ bytes) + a few residual bits ($\beta$ bits), which was adopted by SZ \cite{sz16,tpds-point-wise}. In this solution, the residual bits with varied number of bits still need to be gathered in a target array by a set of bit-wise operations. 

By comparison, we develop an ultra-fast method (solution C as shown in Figure \ref{fig:cpu-perf-opt}) to deal with the necessary bits very efficiently. The basic idea is bitwise right-shifting the normalized value by $s$ bits, where $s$ is given in Formula (\ref{eq:rightshift}), such that the number of the necessary bits to be stored is always a multiple of 8. As such, the necessary mantissa bits can be represented by an integer number of bytes, with eliminated residual bits. In this situation, we just need to use memory copy operation to commit the necessary bits to one byte-array, which would be fairly fast.  
\begin{equation}
\label{eq:rightshift}
s = \left\{ \begin{array}{l}
 0,\hspace{18mm}R_k\% 8 = 0 \\ 
 8 - R_k\% 8,\hspace{4.5mm}R_k\% 8 \ne 0 \\ 
 \end{array} \right.
\end{equation}

\subsubsection{Investigating Space Overhead of Bitwise Right-Shifting}

Note that the bitwise right-shifting operation may increase the total number of required bits to store, thus reducing the compression ratios in turn. In the following text, we will show that the increased number of bits per value because of the bitwise right-shifting operation is very limited compared with the compressed data size, thanks to the design of identical leading bytes. That being said, such a space overhead introduced to compressed data size is  negligible in most cases. In fact, although the bitwise right-shifting operation may increase the required number of bits, this operation may also potentially increase the number of identical leading bytes, such that some necessary bits could be ``recorded'' by the identical leading array instead. In other words, after the bitwise right-shifting operation, the necessary bits tend to increase on the right end but tend to decrease on its left end, which forms a counteraction to a certain extent. 

\begin{figure}[ht] \centering

\hspace{-8mm}
\subfigure[{Hurricane-ISABEL ($e$=1E-3)}]
{
\raisebox{-1cm}{\includegraphics[scale=0.39]{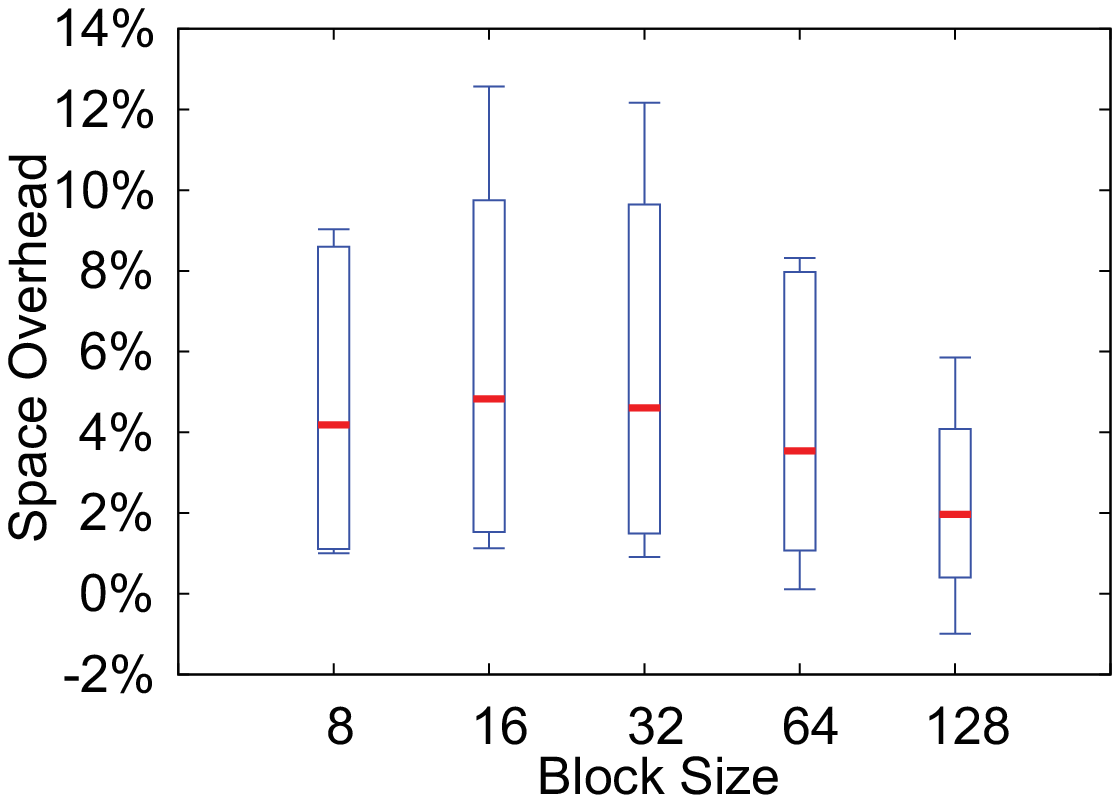}}
}
\hspace{-7mm}
\subfigure[{Miranda ($e$=1E-3)}]
{
\raisebox{-1cm}{\includegraphics[scale=0.39]{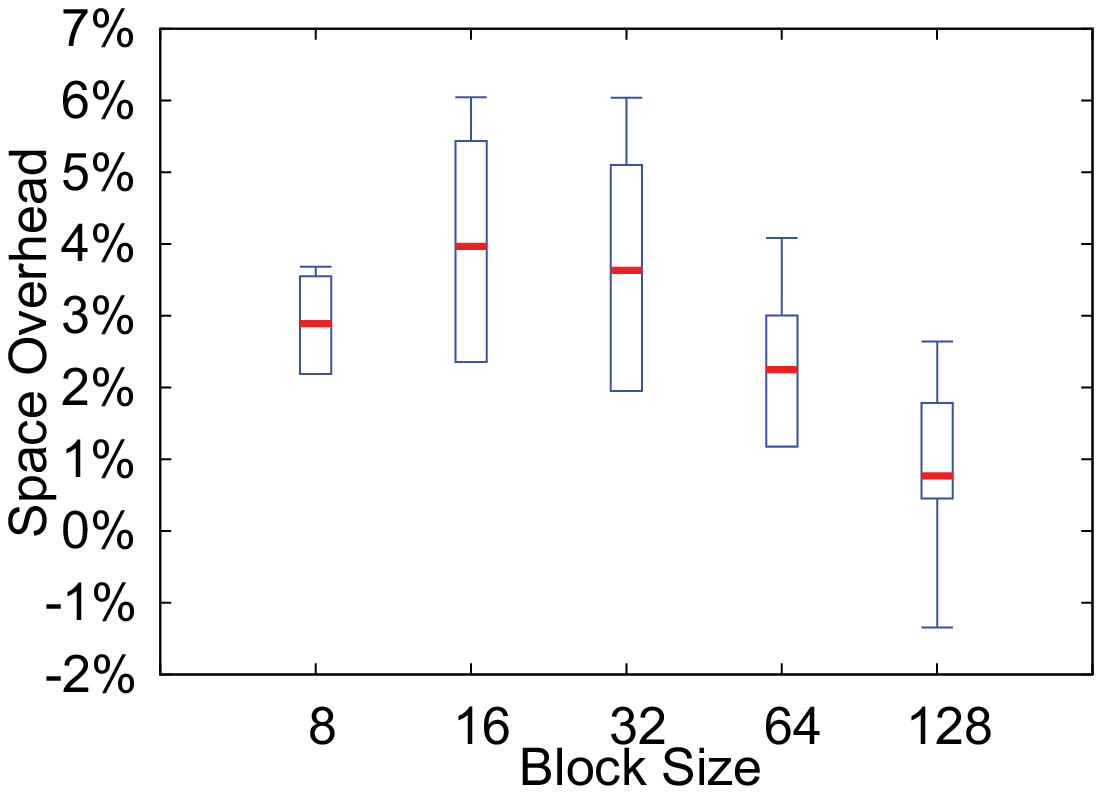}}
}
\hspace{-10mm}

\hspace{-8mm}
\subfigure[{Hurricane-ISABEL ($e$=1E-4)}]
{
\raisebox{-1cm}{\includegraphics[scale=0.39]{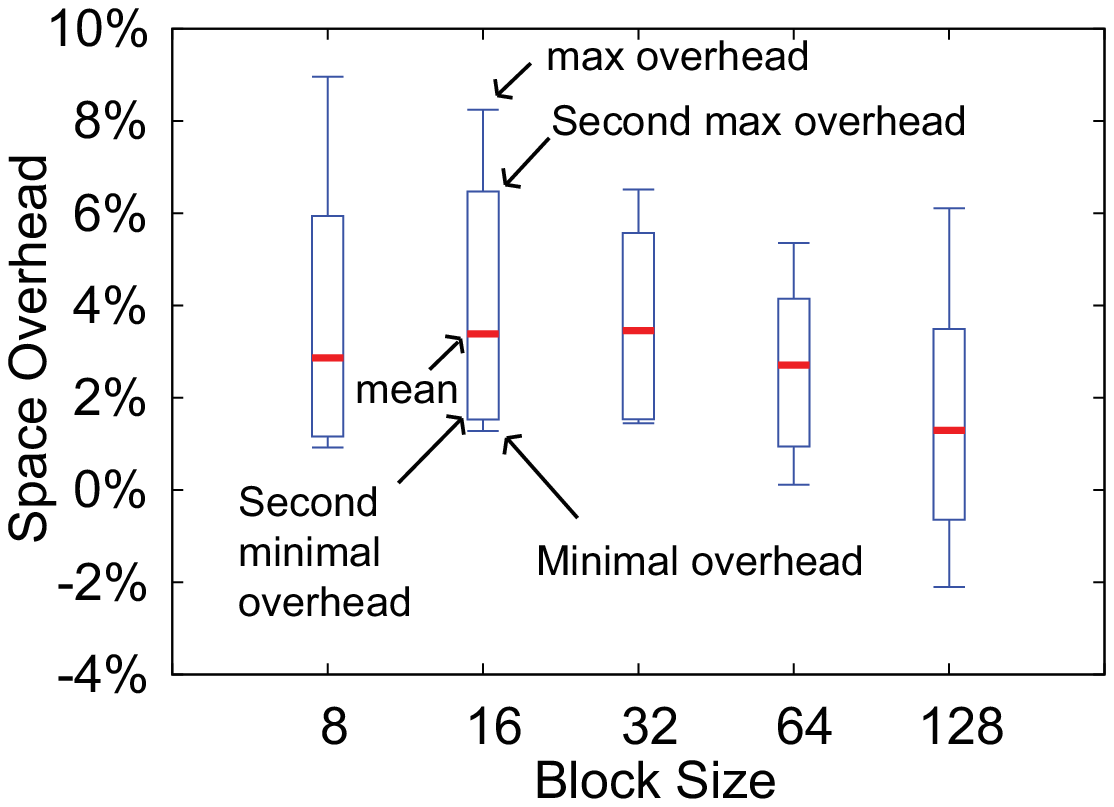}}
}
\hspace{-6mm}
\subfigure[{Miranda ($e$=1E-4)}]
{
\raisebox{-1cm}{\includegraphics[scale=0.39]{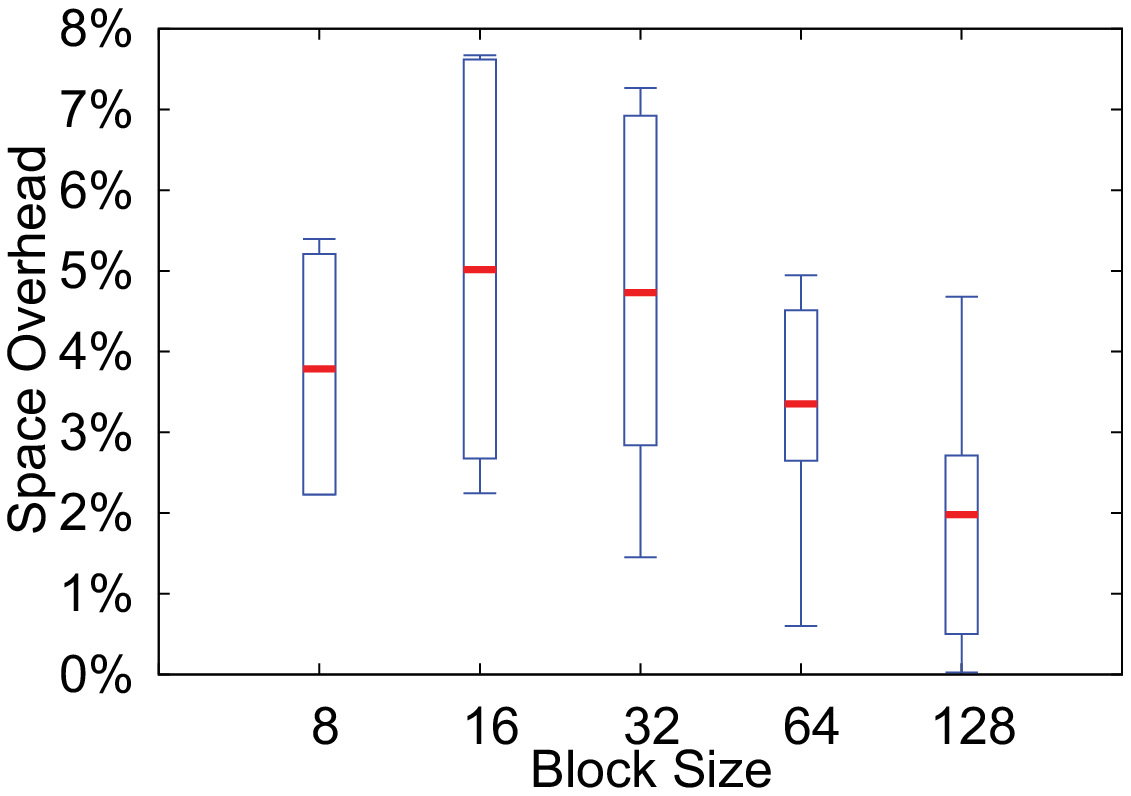}}
}
\hspace{-10mm}
\vspace{-1mm}

\hspace{-8mm}
\subfigure[{Hurricane-ISABEL ($e$=1E-5)}]
{
\raisebox{-1cm}{\includegraphics[scale=0.39]{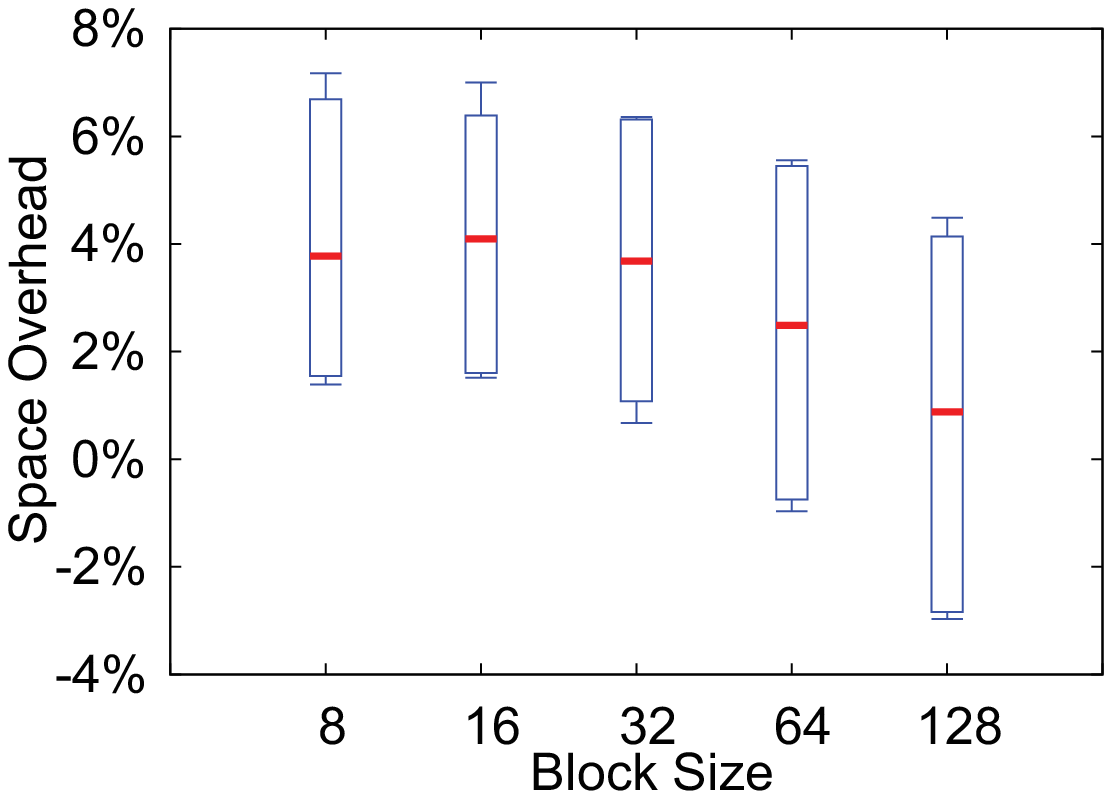}}
}
\hspace{-6mm}
\subfigure[{Miranda ($e$=1E-5)}]
{
\raisebox{-1cm}{\includegraphics[scale=0.39]{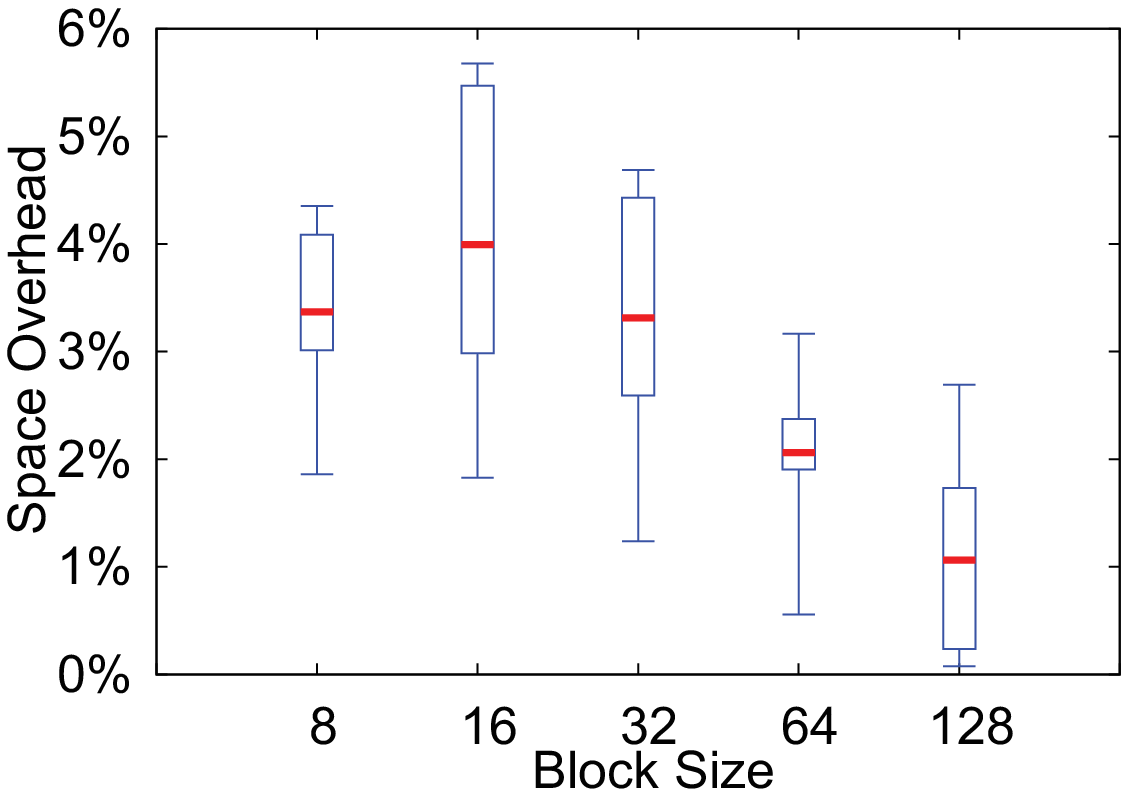}}
}
\hspace{-10mm}
\vspace{-1mm}

\caption{Space overhead of bitwise right-shifting used in {\projectName}, showing the min, 2nd-min, avg, 2nd-max, and max overhead for two application datasets each with multiple fields}
\label{fig:space-overhead}
\end{figure}

We use Figure \ref{fig:space-overhead} (based on two real-world simulation datasets with different value range based error bounds \cite{z-checker}) to show the specific space overhead of our solution designed with bitwise right-shifting operation, as compared with the compressed data size. Specifically, the space overhead is defined as the ratio of the increased storage space introduced by the bitwise right-shifting method to the compressed data size, as presented in Formula (\ref{eq:space-overhead}).  
\begin{equation}
\label{eq:space-overhead}
Overhead = \frac{\sum\limits_{v_i\in D}{(R_k+s-L_i')}-\sum\limits_{v_i\in D}{(R_k-L_i)}}{D_{size}/C\hspace{-0.4mm}R}
\end{equation}
where CR is compression ratio, $D_{size}$ refers to the original data size (thus $D_{size}$/CR means compressed data size), $\sum\limits_{v_i\in D}{(R_k+s-L_i')}$ refers to the total amount of necessary bytes to store under the Solution C (our solution), and $\sum\limits_{v_i\in D}{(R_k-L_i)}$ refers to the total amount of necessary bytes to store by Solution A or B.

According to Figure \ref{fig:space-overhead} which involves a total of about 100 different fields across these two applications, it is clearly observed that the space overhead is always lower than 12\% for all the fields, and average overhead for each case (with a specific block size) is always around or lower than 5\% compared with the compressed data size. We give an example to further explain how small the overhead is as follows. Specifically, for the field `density' in the Miranda simulation dataset, the original data size is 256$\times$384$\times$384$\times$bytes = 144MB, the compression ratio of {\projectName} is 9.923, so the compressed data size is about 15.2MB. Our characterization shows that Solution B and Solution C lead to 81,340,334 necessary bits (i.e., 10,167,542 bytes) and 83,054,120 necessary bits (i.e., 10,381,765 bytes), respectively, which means the overhead is only $\frac{10,381,765-10,167,542}{15.2MB}$=1.4\% for this field. 

%Another interesting observation based on Figure \ref{fig:space-overhead} is that the space overhead generally tends to be small for relatively small (such as 8) or relatively big (such as 128) block sizes, and it turns out to be slightly higher when the block size is set to a median value (such as 16 or 32).

\subsubsection{Exploring The Optimal Block Size}

Different block sizes may affect the compressed data sizes (i.e., compression ratios) significantly, thus it is necessary to investigate the most appropriate setting about block size for the {\projectName}. As described previously, there are two types of the blocks in the design, which are called `constant' blocks (line 4$\sim$5 in Algorithm \ref{alg:design-skeleton}) and `nonconstant' blocks (line 6$\sim$12 in Algorithm \ref{alg:design-skeleton}), respectively. Before exploring the optimal block size, we need to understand how the two types of blocks contribute to the compressed data size (or compression ratios), which are analyzed as follows (with three \textit{impact factor}s summarized).

\begin{itemize}
    \item \textit{Analysis of Constant Blocks} Constant blocks refer to the blocks each of which can be approximated by using one data value $\mu_k$ (i.e., mean of min and max). As such, the smaller block size, the more data points to be included in the constant blocks, because of the finer-grained block-wise processing, as illustrated in Figure \ref{fig:constant-block} (a). As shown in the figure, the first set of 8 data points can form a constant block because of relatively small block size. In this sense, the compression ratio tend to increase with decreasing block size because all the values within the constant block can be approximated by one value (i.e., $\mu_k$), which is called \textbf{impact factor} \circled{A} in the following text.  
    However, since each constant block needs to store a constant value $\mu_k$ in the compressed data, the smaller block size, the larger number of $\mu_k$ to be stored, which may also decrease the compression ratio in turn, as illustrated in Figure \ref{fig:constant-block} (b). We call this phenomenon \textbf{impact factor} \circled{B}. Specifically, for the relatively smooth regions in the dataset, the algorithm still needs to store multiple $\mu_k$s even though a large number of adjacent data points could be approximated by only one uniform value instead. This may introduce significant overhead because of extra unnecessary $\mu_k$ to store, thus leading to the lower compression ratios.
\begin{figure}[ht] \centering

\subfigure[{Pros of Small Block Size}]
{
\raisebox{-1cm}{\includegraphics[scale=0.55]{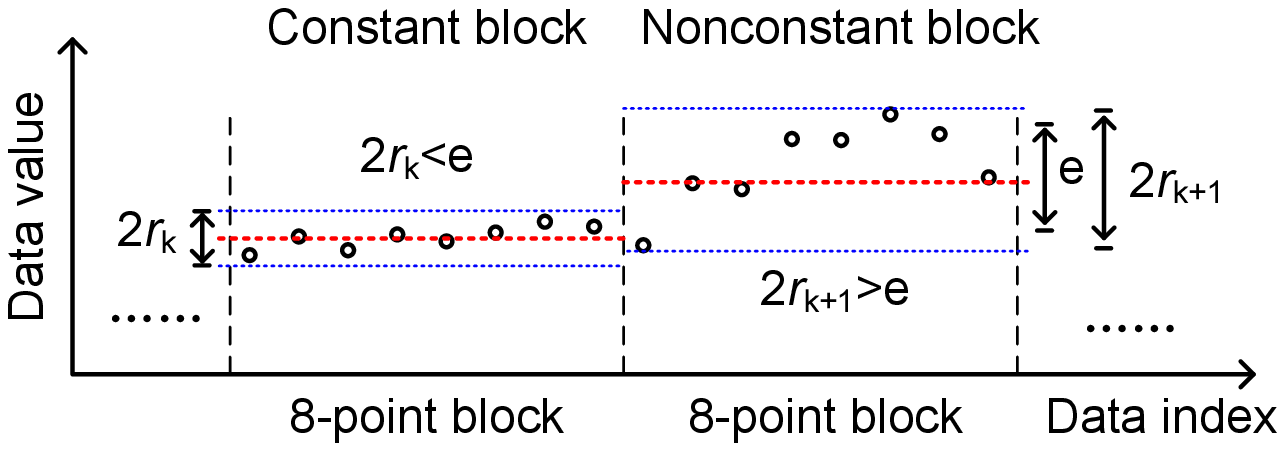}}
}

\hspace{1mm}\subfigure[{Cons of Small Block Size}]
{
\raisebox{-1cm}{\includegraphics[scale=0.55]{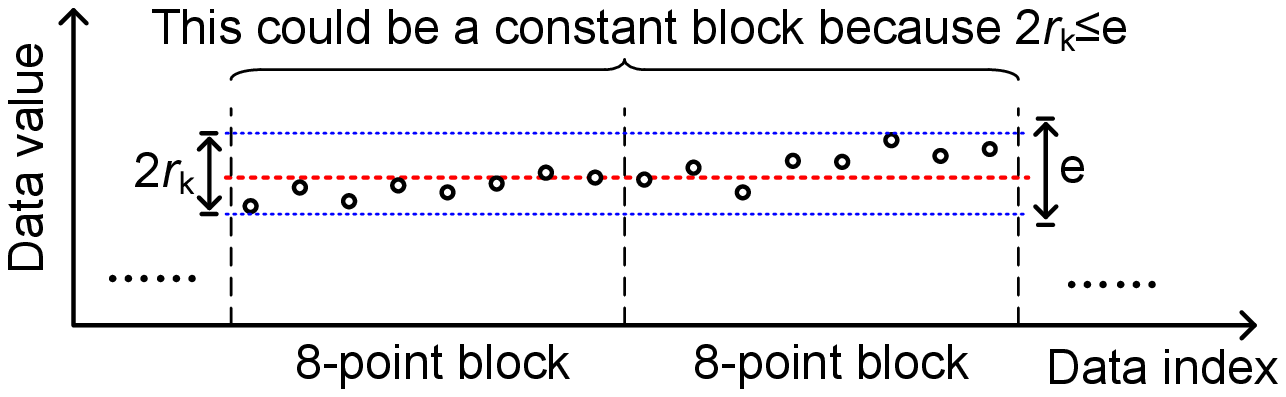}}
}
\vspace{-2mm}
\caption{Constant block's pros and cons when block size is small}
\label{fig:constant-block}
\end{figure}       
    \item \textit{Analysis of Nonconstant Blocks} On the one hand, the impact factor \circled{B} also applies on nonconstant blocks as they also need to store $\mu$ for data denormalization during the decompression. On the other hand, smaller block size may tend to get higher compression ratios, because of the following reason (we call it \textbf{impact factor} \circled{C}). In fact, the smaller block size, the smaller variation in the block (i.e., smaller $\mu_k$), and thus the fewer necessary bits to store. Specifically, as shown in Figure \ref{fig:constant-block} (a), the first 8-point block has much smaller data variation than the other one, so that the corresponding required exponent would be smaller, leading to fewer required mantissa bits (according to Formula (\ref{eq:required-bits})).  
\end{itemize}

\begin{figure}[ht] \centering

\hspace{-8mm}
\subfigure[{CR ($e$=1E-3)}]
{
\raisebox{-1cm}{\includegraphics[scale=0.38]{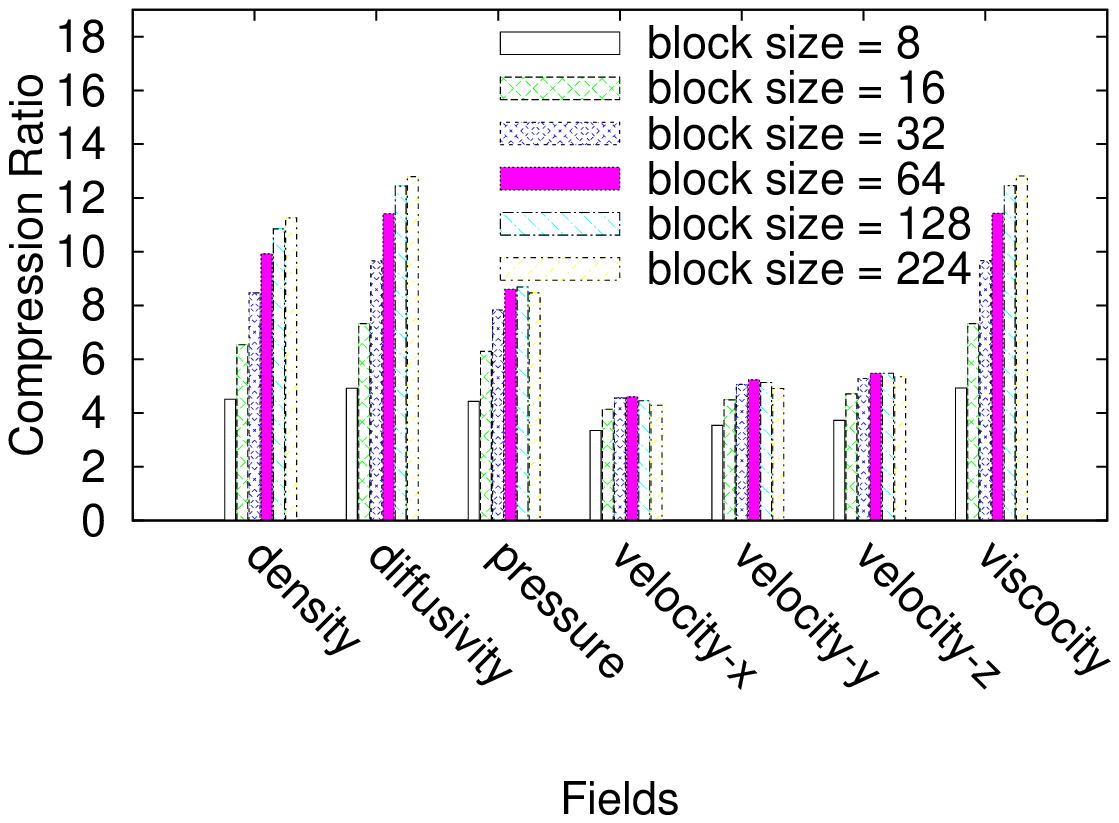}}
}
\hspace{-10mm}
\subfigure[{CR ($e$=1E-4)}]
{
\raisebox{-1cm}{\includegraphics[scale=0.38]{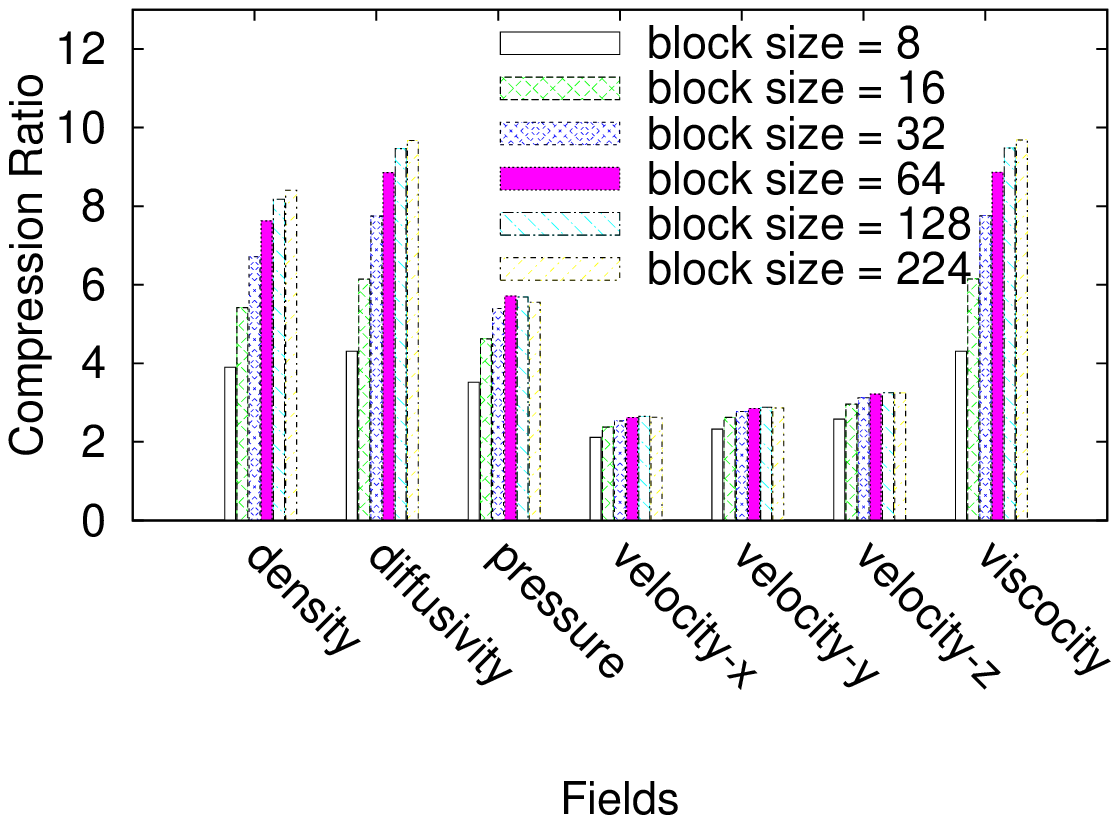}}
}
\hspace{-10mm}
\vspace{-1mm}

\hspace{-8mm}
\subfigure[{PSNR ($e$=1E-3)}]
{
\raisebox{-1cm}{\includegraphics[scale=0.38]{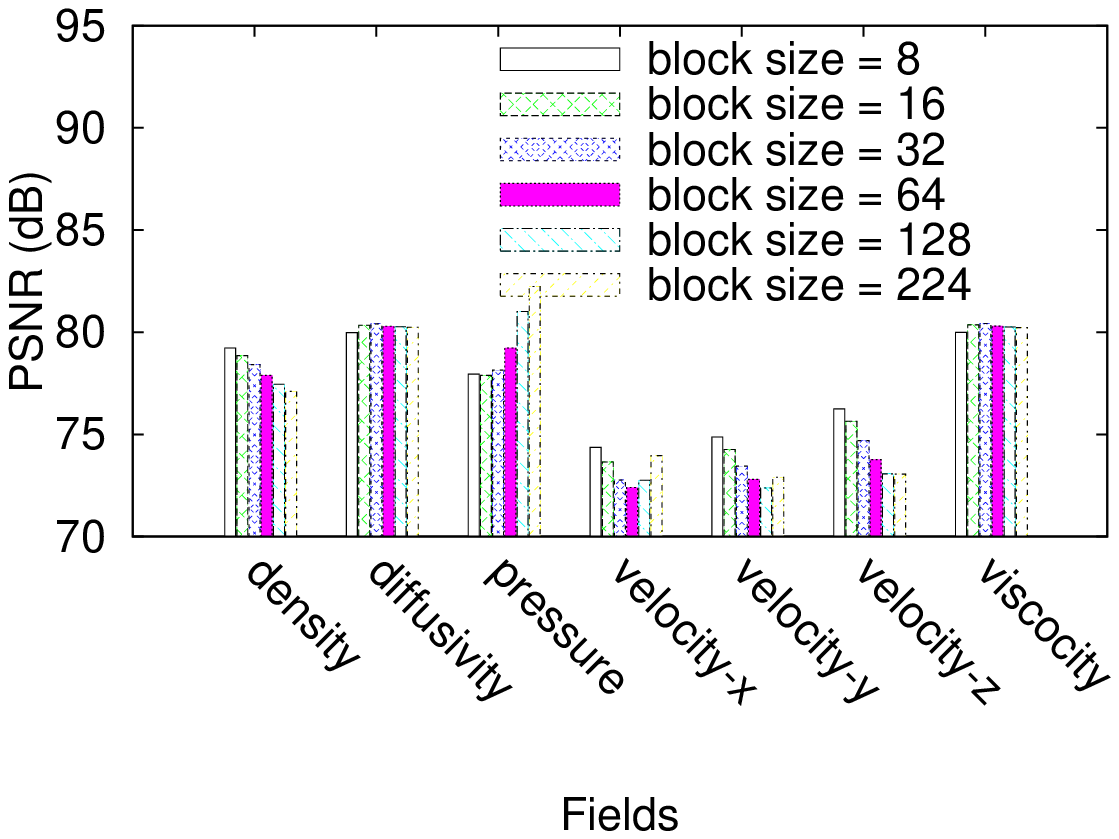}}
}
\hspace{-10mm}
\subfigure[{PSNR ($e$=1E-4)}]
{
\raisebox{-1cm}{\includegraphics[scale=0.38]{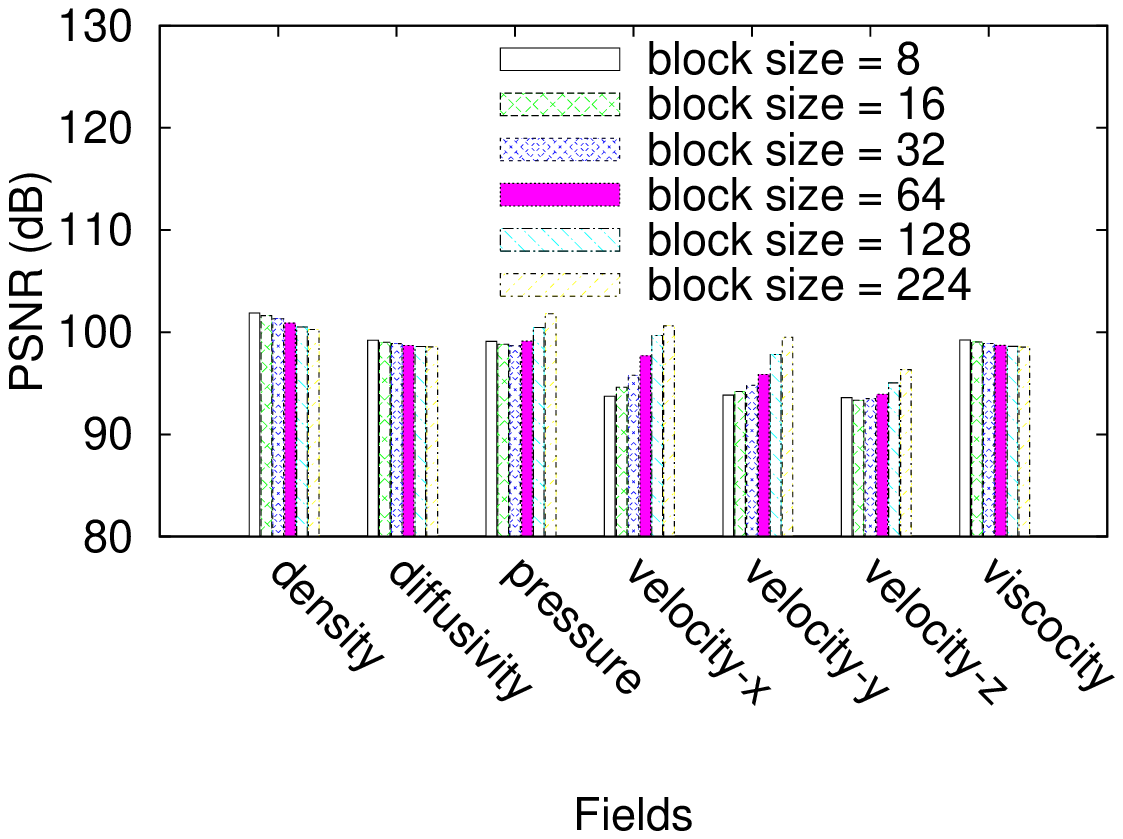}}
}
\hspace{-10mm}
\vspace{-1mm}

\caption{Compression quality of Miranda data with various block sizes}
\label{fig:block-ratio-psnr}
\end{figure}

Based on the above analysis, different block sizes may have distinct pros and cons to the compression quality in the regard of the two different types of blocks. It is not obvious what block size can get the best compression quality. In what follows, we explore the best block size setting by characterizing the compression ratios and Peak Signal to Noise Ratio (PSNR) with different block sizes, as presented in Figure \ref{fig:block-ratio-psnr}. PSNR is a critical lossy compression data quality assessment metric, which has been widely used in the lossy compression and visualization community \cite{z-checker, sz16,sz17,szauto,zfp}. PSNR is defined in Formula (\ref{eq:psnr}). 
\begin{equation}
\label{eq:psnr}
psnr = 20\log_{10}\frac{(d_{\max}-d_{\min})}{\sqrt{MSE}}
\end{equation}
where $d_{min}$ and $d_{max}$ are the min value and max value in the dataset $D$, and MSE refers to the mean squared error between the original dataset $D$ and reconstructed dataset $D'$. The higher PSNR, the higher precision of the reconstructed data.

In the exploration, we checked many different error bounds from 1E-3 through 1E-6. Because of space limit, we present in Figure \ref{fig:block-ratio-psnr} only the results about the value range based error bound of 1E-3 and 1E-4, which compress 7 fields of the Miranda simulation dataset by {\projectName}. Other error bounds and datasets exhibit very similar results. 

From Figure \ref{fig:block-ratio-psnr}, we can observe that the compression ratio increases with block size in most of cases, while the PSNR always stays at the same level across different block size settings. This motivates us the best block size is 128 based on such a comprehensive characterization covering multiple applications and error bounds. This characterization also indicates that the \textbf{impact factor} \circled{B} dominates the overall compression ratios, because this is the only factor that may enhance the compression ratio with increasing block size. 

\subsection{Performance Optimization for GPU}
In this section, we describe our design and implementation for the cu{\projectName} - CUDA GPU version of {\projectName}. {\projectName} is 
considered as an irregular application to the GPU platforms due to the data dependencies 
in both the compression and decompression. In order to maximize the utilization of 
the GPU computing capacity, algorithmic adjustments and architecture-specific optimizations 
for the cu{\projectName} are needed.

\noindent\textbf{Compression}: The basic design of cu{\projectName}'s compression is that each CUDA thread-block
handles one data-block. A thread-block is configured in two-dimensions. The data-block size is 
chosen as a multiple of warp size to optimize the performance. There are two phases during 
the cu{\projectName} compression. In the first phase, the compressor distinguishes the constant and 
non-constant data-blocks by calculating their $\mu$ values and deviation radius. The entire data-block
set is processed by the thread-blocks iteratively. A thread-block with a non-constant 
data-block would enter the second phase, whereas the one with a constant data-block would 
record the data-block index and then immediately go forward to process the next data-block. 
With this fashion, the workload imbalance can be significantly mitigated, in that the number of data-blocks is considerably larger than the number of thread-blocks. We also note that the calculations of min and max value are the main computing workloads during the first phase, and they can be efficiently 
parallelized by leveraging the CUDA warp-level operations.

Only the thread-blocks with non-constant data-blocks go through the second phase. In the 
second phase, the leading-number-based compression is performed. Performing the computation of \emph{xor\_leadingzero\_array} 
and \emph{mb\_array} (a.k.a., mid-bytes) on GPU is straightforward. However, writing the mid-bytes back 
to global memory in a compact format is challenging. Unlike the serial implementation, 
in cu{\projectName}, the starting address to write mid-bytes of each data-block remains unknown
until the total number of mid-bytes of all its preceding data-blocks is computed. Therefore,
a prefix scan should be performed before writing the mid-bytes to the memory. Prefix-scan 
on GPU has been well studied~\cite{scan}. We leverage the classical design and implement 
it using 2-level in-warp shuffles.

\noindent\textbf{Decompression}: The basic design of cu{\projectName}'s decompression is similar to 
its compression, in which each thread-block handles one data-block. Since the decompression 
of the constant data-block is very lightweight, we only decompress the non-constant 
data-blocks in GPU. The decompression consists of two components: the leading-byte retrieval 
and the mid-byte retrieval. Implementing the latter is relatively straightforward as it 
just need to read the bytes from the compressed data. We note that, due to the same 
reason as in the compression, a parallel prefix-scan is applied before retrieving the 
mid-bytes. However, implementing the leading-byte retrieval is challenging since retrieving 
the bytes from the preceding adjacent element no longer works in the parallel environment.

\begin{figure}[ht]
    \centering
    \includegraphics[width=0.95\columnwidth]{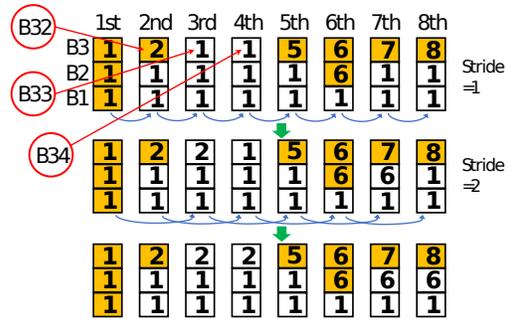}
    \vspace{-2mm}
    \caption{Schematic graph showing the index propagation for  parallel leading-byte retrievals (assume the number of required bytes for this block is 3).}
    \label{fig:idx-prop}
\end{figure}

In Figure \ref{fig:idx-prop}, we illustrate the challenge by an example (see first row in particular), which displays an eight-element data-block with the mid-bytes highlighted in yellow.
In the serial code, we can retrieve the third leading-byte of the third element (B33) by reading
the mid-byte of the second element (B32) and then subsequently retrieve B34 by reading B33. However, 
in the parallel context, B33 and B34 will be simultaneously retrieved, which will cause the read-after-write (RAW)
hazard. The same issue occurs when B27 and B28 are retrieved simultaneously. 

To solve the above issue, 
the decompression needs to predetermine the mid-byte position each leading-byte should read from.
For example, in this case, both B33 and B34 should read from B32. To this end, we propose 
an index-propagation. This algorithm assigns each byte an initial reading-position. The ones for 
leading-bytes are the first element's index (1 in the example) while the ones for mid-bytes are 
their own elements' indices. Then a parallel interleaved addressing propagation will be 
performed to propagate the position values. If the source value greater than the destination 
value, the source value will be used as the new reading position for the current byte. With 
the interleaved addressing scheme, the propagation complexity is reduced to O(logn). In the running 
example, we only need three rounds of shuffles with strides 1, 2, and 4 to finish the propagation.
Notice that we omit the shuffles with stride=4 in Figure \ref{fig:idx-prop} to save space. It does
not change the final position values in this case. After the index-propagation, each leading-byte
knows which mid-byte it should read from to retrieve its value as shown in the last row of Figure \ref{fig:idx-prop}.

\section{Performance Evaluation}
\label{sec:evaluation}

In this section, we analyze the evaluation results, which are performed using 6 real-world scientific datasets on heterogeneous devices offered by two different supercomputers.

\subsection{Experimental Setup}

Table \ref{tab:app} describes all the application datasets used in our experiments. All the datasets are downloaded from the well-known scientific data reduction benchmark website \cite{sdrbench}. 

\begin{table*}[h]
\centering
\caption{Applications (all datasets here are originally stored in single-precision floating-points)}
\vspace{-2mm}
\begin{tabular}{|c|c|c|l|}
\hline
\textbf{Application} & \textbf{\# fields} &\textbf{Size per field} &\textbf{Description} \\ \hline
CESM-ATM (CE.) \cite{cesm} & 77 &1800$\times$3600 & Atomosphere simulation of Community Earth System Model\\ \hline
Hurricane (Hu.) \cite{hurricane-2004} & 13 &100$\times$500$\times$500 & simulation of Hurricane ISABEL \\ \hline
Miranda (Mi.) \cite{Miranda} & 7 & 256$\times$384$\times$384 &large-eddy simulation of multicomponent flows with turbulent mixing \\ \hline
Nyx (Ny.) \cite{nyx} & 6 & 512$\times$512$\times$512 & adaptive mesh, massively-parallel, cosmological simulation \\ \hline
QMCPack (QM.) \cite{qmcpack} & 2 & 288/816$\times$115$\times$69$\times$69 & simulation for electronic structure of atoms, molecules and solids \\ \hline
SCALE-LetKF (SL.) \cite{scale-letkf} & 12 & 98$\times$1200$\times$1200 & SCALE-RM weather simulation based on LETKF filter \\ \hline
\end{tabular}
\label{tab:app}
\end{table*}

We perform our GPU experiments on both A100 GPU (offered by ANL ThetaGPU \cite{thetaGPU}) and V100 GPU (offered by ORNL Summit \cite{summit}). We perform our CPU experiments on ANL ThetaGPU's compute nodes. We compare our developed ultra-fast compressor {\projectName} with two outstanding lossy compressors -- SZ \cite{sz16,sz17} and ZFP \cite{zfp}, since they are arguably the fastest existing error-bounded compressors based on prior studies \cite{szauto,sz16} and they both have GPU versions that can be compared with our solution {\projectName} in the experiments. 

\subsection{Evaluation Results}

\begin{figure}[ht] 
\centering
\hspace{-12mm}
\subfigure[{original data}]
{
\raisebox{-1cm}{\includegraphics[scale=0.26]{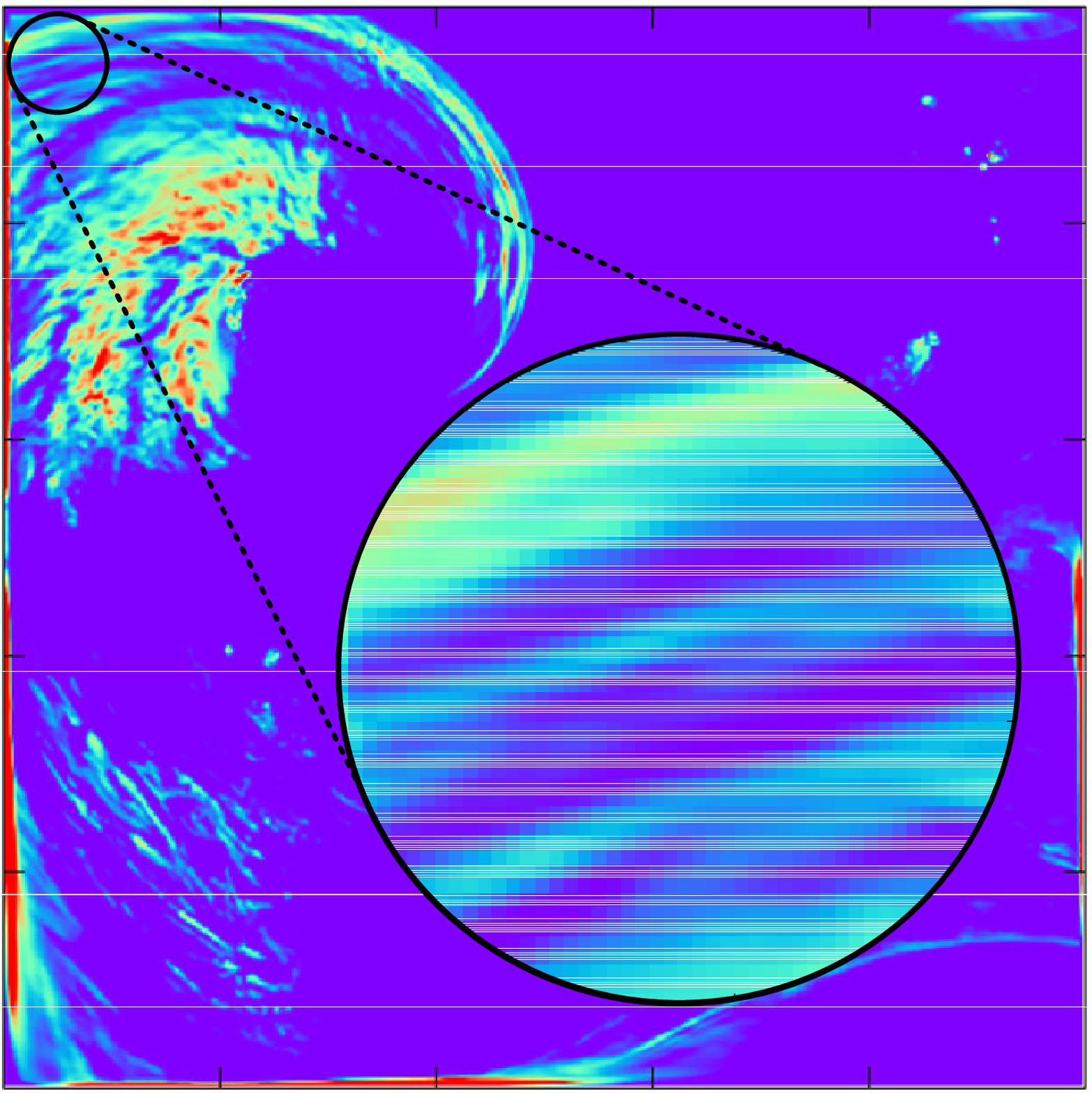}}
}
\hspace{-4mm}
\subfigure[{$e$=1E-3,PSNR=74.4,SSIM=0.93}]
{
\raisebox{-1cm}{\includegraphics[scale=0.26]{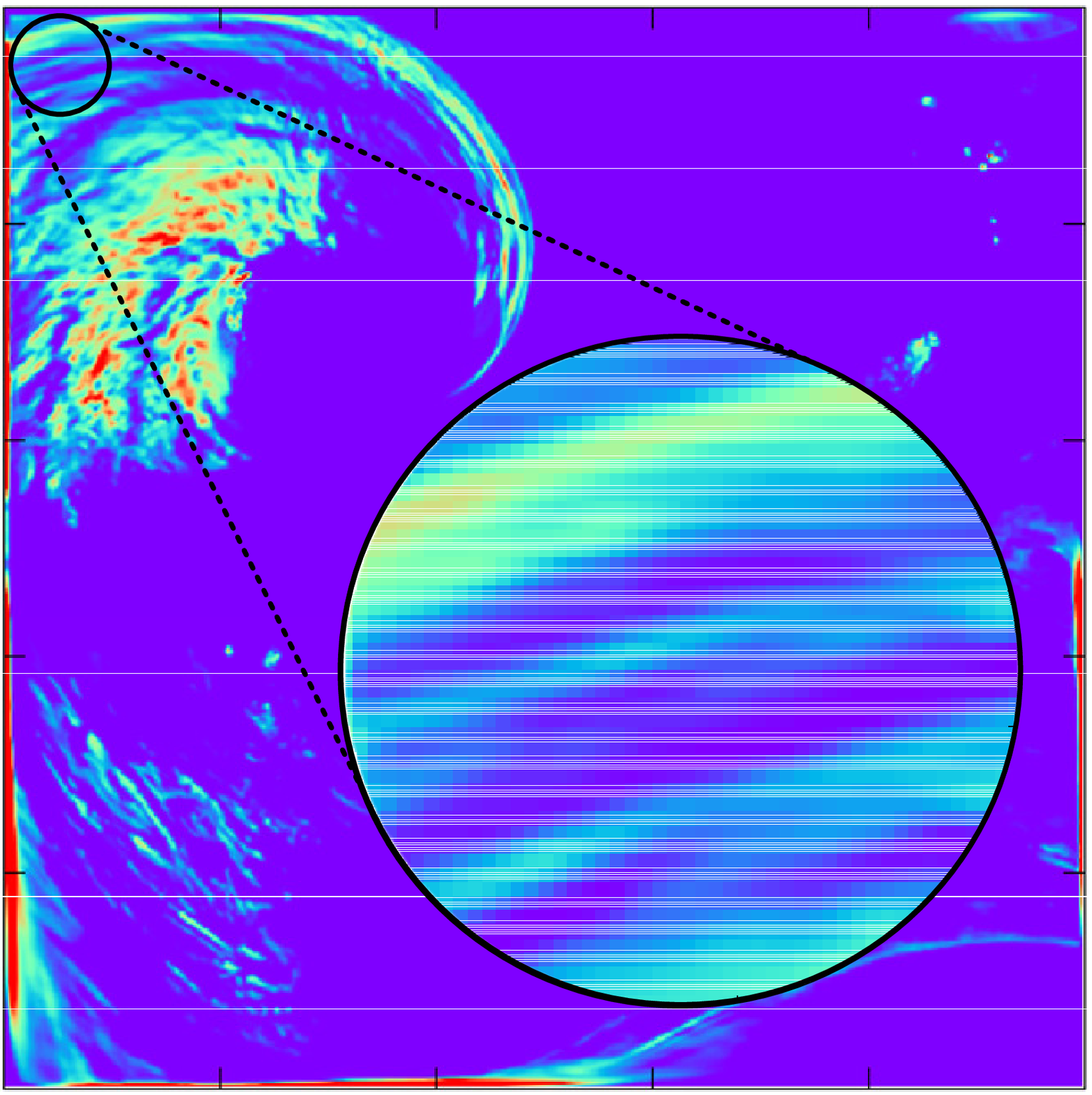}}
}
\hspace{-12mm}
\vspace{-1mm}

\hspace{-12mm}
\subfigure[{$e$=4E-3,PSNR=62,SSIM=0.89}]
{
\raisebox{-1cm}{\includegraphics[scale=0.26]{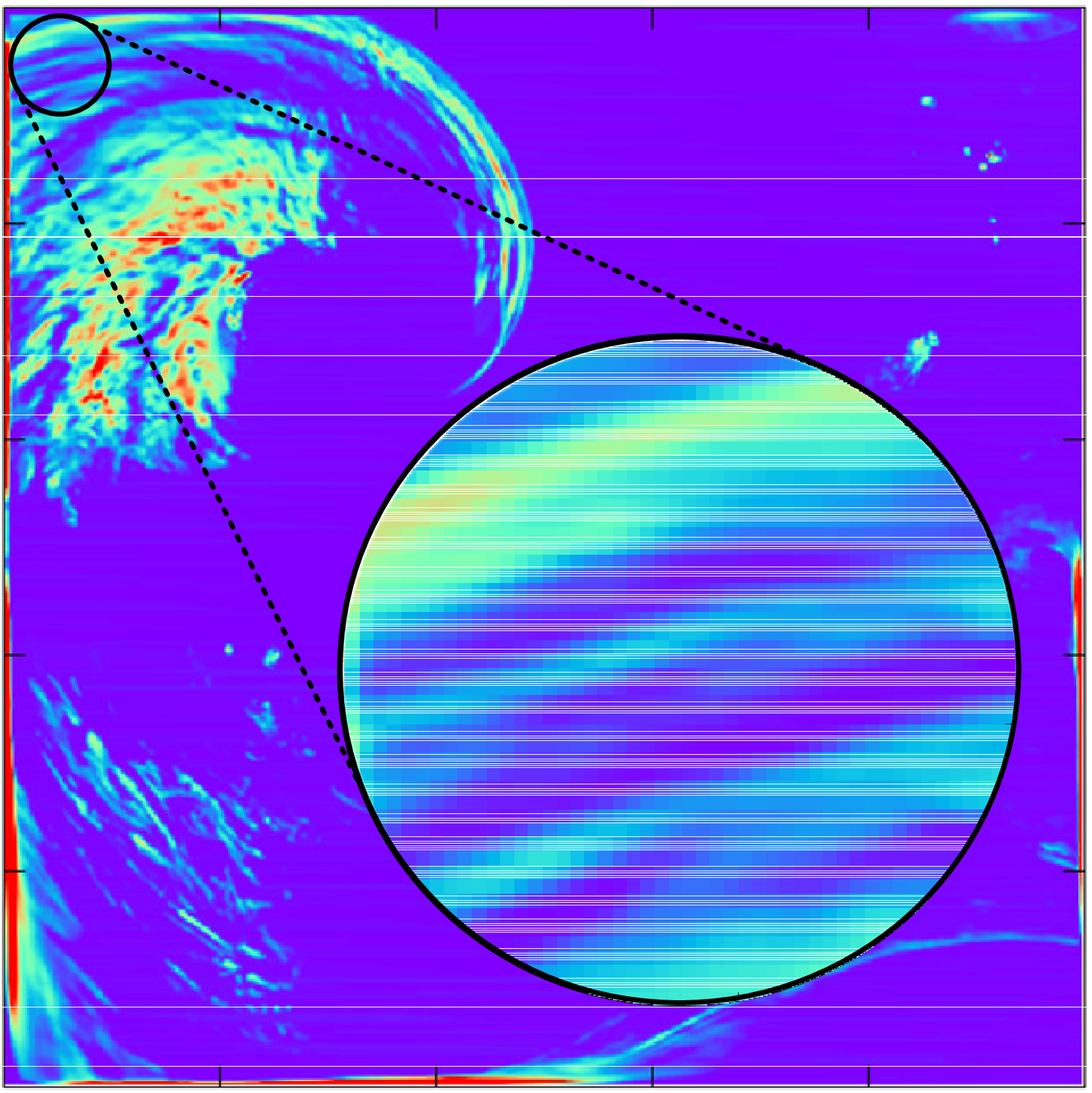}}
}
\hspace{-4mm}
\subfigure[{$e$=1E-2,PSNR=54.6,SSIM=0.865}]
{
\raisebox{-1cm}{\includegraphics[scale=0.26]{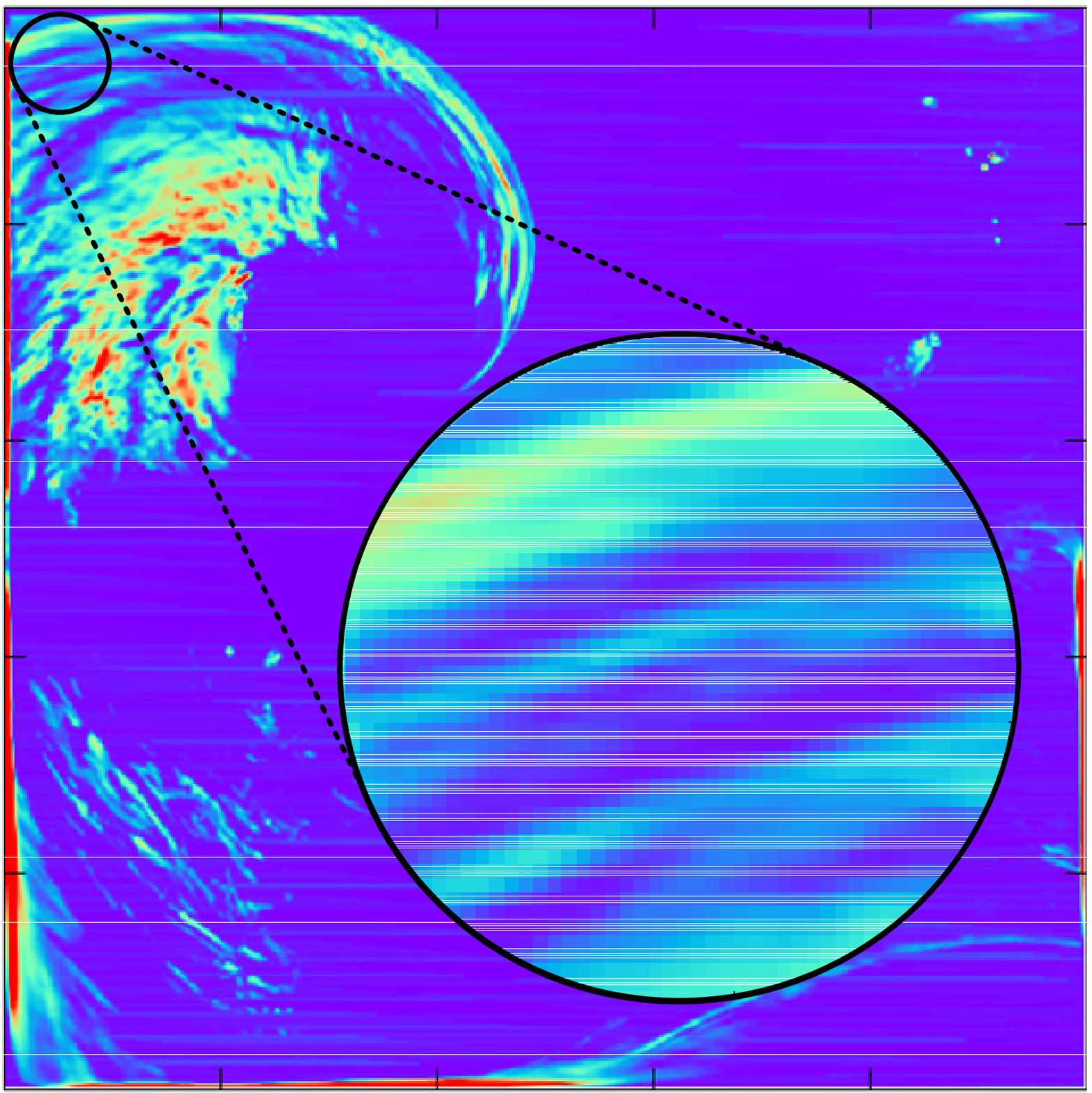}}
}
\hspace{-12mm}
\vspace{-1mm}

\caption{Visual Quality of {\projectName} on Hurricane-ISABEL Simulation (the compression ratios are 14.6, 18, 20.64, respectively)}
\label{fig:visualization-hurricane}
\end{figure}

First of all, we check the data reconstruction quality under our {\projectName} for all the simulation datasets involved in our experiments. We conclude that the overall visual quality looks great when the value range based error bound (denoted by $REL$) is set to 1E-2$\sim$1E-4 for {\projectName}. Due to space limit, we demonstrate the visual quality, PSNR, and Structural Similarity Index (SSIM) using only the Hurricane-SIABEL simulation dataset (CLOUDf48), as shown in Figure \ref{fig:visualization-hurricane} (compression ratios are 14.6, 18, and 20.64, respectively). We can observe that the reconstructed data's visual quality is pretty high, even zooming in the top-left corn by 50$\times$, though a few artifacts can be seen in the dark blue area of Figure \ref{fig:visualization-hurricane} (d). How to further mitigate or remove artifacts will be our future work. 

We present compression ratios (CR) of the three compressors in Table \ref{tab:cr}, by showing the minimal, overall (i.e., Harmonic mean), and maximum CR, respectively, for all the fields in each application. The table shows that {\projectName} can get very high compression ratios (e.g., 124 for CESM) when REL=1E-2. Its overall compression ratio is 3$\sim$12 in all the cases, which is 0.5$\sim$3$\times$ lower compared with ZFP and 3$\sim$30$\times$ lower compared with SZ. This is reasonable because SZ and ZFP adopt advanced multidimensional data analysis and sophisticated encoding methods, which can get fairly high compression ratios but may suffer lower execution performance on both CPU and GPU in turn (to be shown later). By comparison, the overall compression ratio for the lossless compressor Zstd is only 1.12$\sim$1.49, which is lower than that of {\projectName} by about 200$\sim$400\%.    

\begin{table*}[ht]
\centering
\footnotesize
\caption{Compression Ratios (Original Data Size / Compressed Data Size)}
\vspace{-2mm}
\begin{adjustbox}{max width=0.99\textwidth}
\begin{tabular}{|c|c||c|c|c|c|c|c|c|c|c|c|c|c|c|c|c|c|c|c|}
\hline
\textbf{} &  & \multicolumn{3}{c|}{\textbf{CESM}} & \multicolumn{3}{c|}{\textbf{Hurricane}} & \multicolumn{3}{c|}{\textbf{Miranda}} & \multicolumn{3}{c|}{\textbf{Nyx}} & \multicolumn{3}{c|}{\textbf{QMCPack}} & \multicolumn{3}{c|}{\textbf{Scale-LetKF}} \\ \hline
 & \textbf{REL} & min & avg & max & min & avg & max & min & avg & max & min & avg & max & min & avg & max & min & avg & max  \\ \hline
 & 1E-2 & 4 & 9.1 & 124 & 4 & 6.6 & 21.1 & 8.2 & 11.8 & 16.2 & 4.8 & 11.34 & 124 & 9.2 & 9.4 & 9.7 & 7.6 & 10.6 & 25.2  \\ \cline{2-20}
\textbf{\projectName} & 1E-3 & 2.84 & 4.61 & 19.3 & 2.9 & 4 & 17.6 & 4.5 & 7.2 & 12.5 & 3.2 & 5.9 & 119 & 4.3 & 4.4 & 4.4 & 3.65 & 4.7 & 7.8   \\ \cline{2-20}
 & 1E-4 & 2.14 & 3.3 & 17 & 2.1 & 3 & 16.2 & 2.7 & 4.5 & 9.5 & 2.4 & 3.7 & 75 & 2.9 & 2.9 & 2.9 & 2.7 & 3.14 & 5.6  \\ \hline\hline
  & 1E-2 & 8 & 13.6 & 46 & 6.4 & 11.3 & 25.8 & 30.5 & 46.6 & 74.6 & 22.5 & 38.8 & 1.1k & 39.1 & 39.2 & 39.4 & 9.4 & 14.5 & 23.8 \\ \cline{2-20}
\textbf{ZFP} & 1E-3 & 4.3 & 7.9 & 30 & 4.3 & 6.7 & 13.2 & 20.6 & 25.6 & 38.5 & 8.2 & 13.1 & 150 & 21 & 21.1 & 21.2 & 6.4 & 7.8 & 13.4  \\ \cline{2-20}
     & 1E-4 & 3 & 5.1 & 18.8 & 2.9 & 4.32 & 10.4 & 11 & 14.5 & 22.9 & 4.1 & 6.2 & 74 & 10.3 & 10.3 & 10.4 & 3.9 & 4.6 & 7.7  \\ \hline\hline
  & 1E-2 & 34.4 & 151 & 3k & 20.4 & 49.8 & 339 & 92.8 & 126 & 234 & 263 & 507 & 21k & 201 & 213 & 227 & 26.3 & 84 & 746 \\ \cline{2-20}
\textbf{SZ} & 1E-3 & 15.6 & 151 & 840 & 9.24 & 17.5 & 58.8 & 49.6 & 59.5 & 75.2 & 36.7 & 79 & 3.6k & 52 & 54.3 & 56.8 & 18.9 & 26.5 & 140 \\ \cline{2-20}
 & 1E-4 & 6.4 & 38.3 & 104 & 5.6 & 9.8 & 31 & 25.1 & 29.6 & 35 & 10.3 & 18.2 & 621 & 18.9 & 19.2 & 19.6 & 10 & 13.9 & 23.1 \\ \hline\hline
 \textbf{zstd} & - & 1.03 & 1.44 & 17.1 & 1.08 & 1.49 & 19.56 & 1.6 & 1.21 & 4.86 & 1.08 & 1.12 & 1.14 & 1.18 & 1.19 & 1.2 & 1.08 & 1.37 & 2.95 \\ \hline
\end{tabular}
\end{adjustbox}
\label{tab:cr}
\end{table*}

We present the single-CPU compression throughput and decompression throughput in Table \ref{tab:cr} and Table \ref{tab:dt}, respectively. The numbers shown in the tables are the overall performance considering all the fields for each application. Through the tables, we can observe that our developed {\projectName} compressor significantly outperforms the other two error-bounded lossy compressors in both compression speed and decompression speed. In absolute terms, for compression, {\projectName} is about 2.5$\sim$5$\times$ as fast as ZFP, and about 5$\sim$7$\times$ as fast as SZ. For decompression, {\projectName} is about 2$\sim$4$\times$ as fast as both ZFP and SZ. Such a high performance in {\projectName} is mainly attributed to two factors: (1) the super-lightweight skeleton design (Algorithm \ref{alg:design-skeleton}), and (2) bitwise right-shifting strategy proposed in Section \ref{sec:perfopt}.

\begin{table}[ht]
\centering
\footnotesize
\caption{Overall Compression Throughput on CPU (MB/s)}
\vspace{-2mm}
\begin{tabular}{|c|c||c|c|c|c|c|c|}
\hline
\textbf{} & \textbf{REL} & \multicolumn{1}{c|}{\textbf{CE.}} & \multicolumn{1}{c|}{\textbf{Hu.}} & \multicolumn{1}{c|}{\textbf{Mi.}} & \multicolumn{1}{c|}{\textbf{Ny.}} & \multicolumn{1}{c|}{\textbf{QM.}} & \multicolumn{1}{c|}{\textbf{SL.}} \\ \hline
& 1E-2 & 1034 & 796 & 959 & 1087 & 969 & 1032 \\ \cline{2-8}
\textbf{\projectName} & 1E-3 & 822 & 750 & 833 & 877 & 902 & 703 \\ \cline{2-8}
& 1E-4 & 752 & 662 & 807 & 722 & 813 & 663 \\ \hline\hline
& 1E-2 & 392 & 256 & 249 & 418 & 323 & 258 \\ \cline{2-8}
\textbf{ZFP} & 1E-3 & 288 & 213 & 211 & 284 & 275 & 208 \\ \cline{2-8}
& 1E-4 & 234 & 181 & 280 & 226 & 208 & 174 \\ \hline\hline
& 1E-2 & 236 & 193 & 186 & 258 & 205 & 217 \\ \cline{2-8}
\textbf{SZ} & 1E-3 & 170 & 153 & 161 & 229 & 216 & 156 \\ \cline{2-8}
& 1E-4 & 143 & 130 & 139 & 164 & 147 & 124 \\ \hline
\end{tabular}
\label{tab:ct}
\end{table}

\begin{table}[ht]
\centering
\footnotesize
\caption{Overall Decompression Throughput  on CPU (MB/s)}
\vspace{-2mm}
\begin{tabular}{|c|c||c|c|c|c|c|c|}
\hline
\textbf{} & \textbf{REL} & \multicolumn{1}{c|}{\textbf{CE.}} & \multicolumn{1}{c|}{\textbf{Hu.}} & \multicolumn{1}{c|}{\textbf{Mi.}} & \multicolumn{1}{c|}{\textbf{Ny.}} & \multicolumn{1}{c|}{\textbf{QM.}} & \multicolumn{1}{c|}{\textbf{SL.}} \\ \hline
& 1E-2 & 1221 & 1085 & 1950 & 1450 & 1292 & 1408 \\ \cline{2-8}
\textbf{\projectName} & 1E-3 & 1022 & 1006 & 1546 & 1218 & 1083 & 975 \\ \cline{2-8}
& 1E-4 & 925 & 864 & 1319 & 956 & 928 & 886 \\ \hline\hline
& 1E-2 & 485 & 476 & 498 & 732 & 685 & 360 \\ \cline{2-8}
\textbf{ZFP} & 1E-3 & 327 & 371 & 401 & 455 & 524 & 395 \\ \cline{2-8}
& 1E-4 & 246 & 297 & 327 & 333 & 376 & 299 \\ \hline\hline
& 1E-2 & 559 & 451 & 549 & 635 & 588 & 519 \\ \cline{2-8}
\textbf{SZ} & 1E-3 & 381 & 291 & 444 & 534 & 462 & 334 \\ \cline{2-8}
& 1E-4 & 269 & 229 & 392 & 359 & 282 & 236 \\ \hline
\end{tabular}
\label{tab:dt}
\end{table}

\begin{figure}[ht] 
\centering
\hspace{-12mm}
\subfigure[{thetaGPU (A100)}]
{
\raisebox{-1cm}{\includegraphics[scale=0.38]{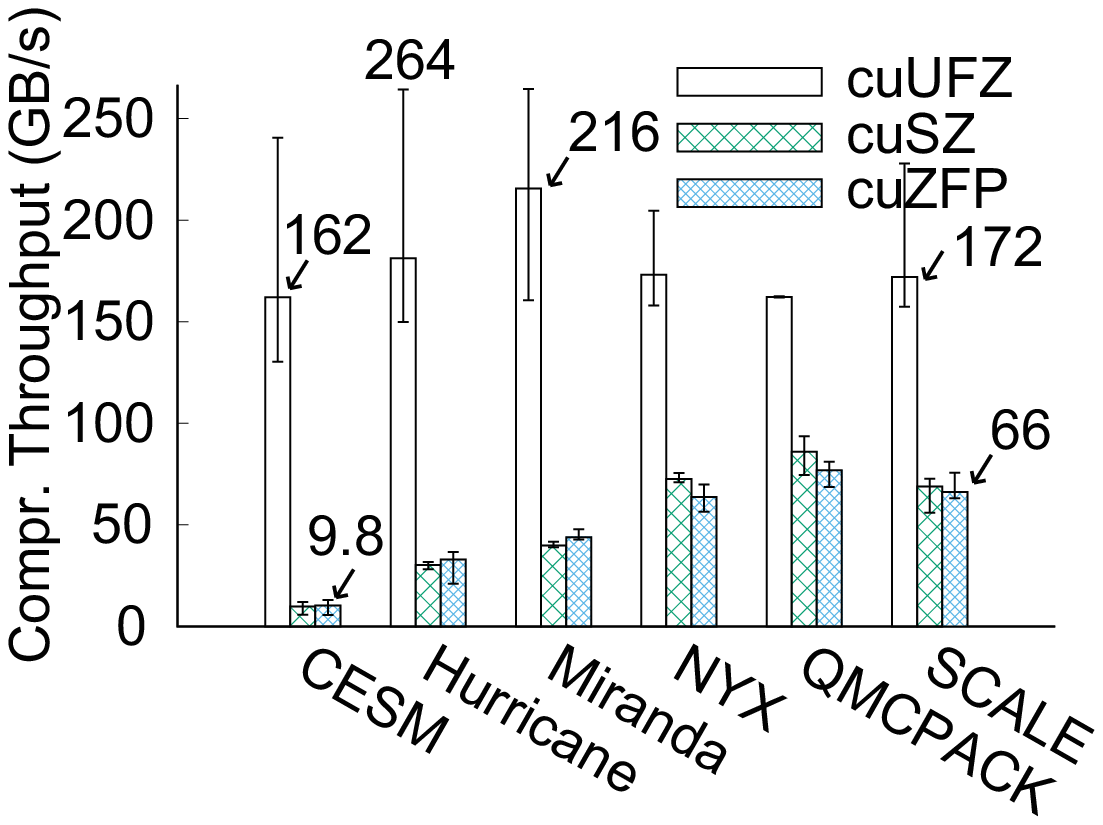}}
}
\hspace{-6mm}
\subfigure[{Summit (V100)}]
{
\raisebox{-1cm}{\includegraphics[scale=0.38]{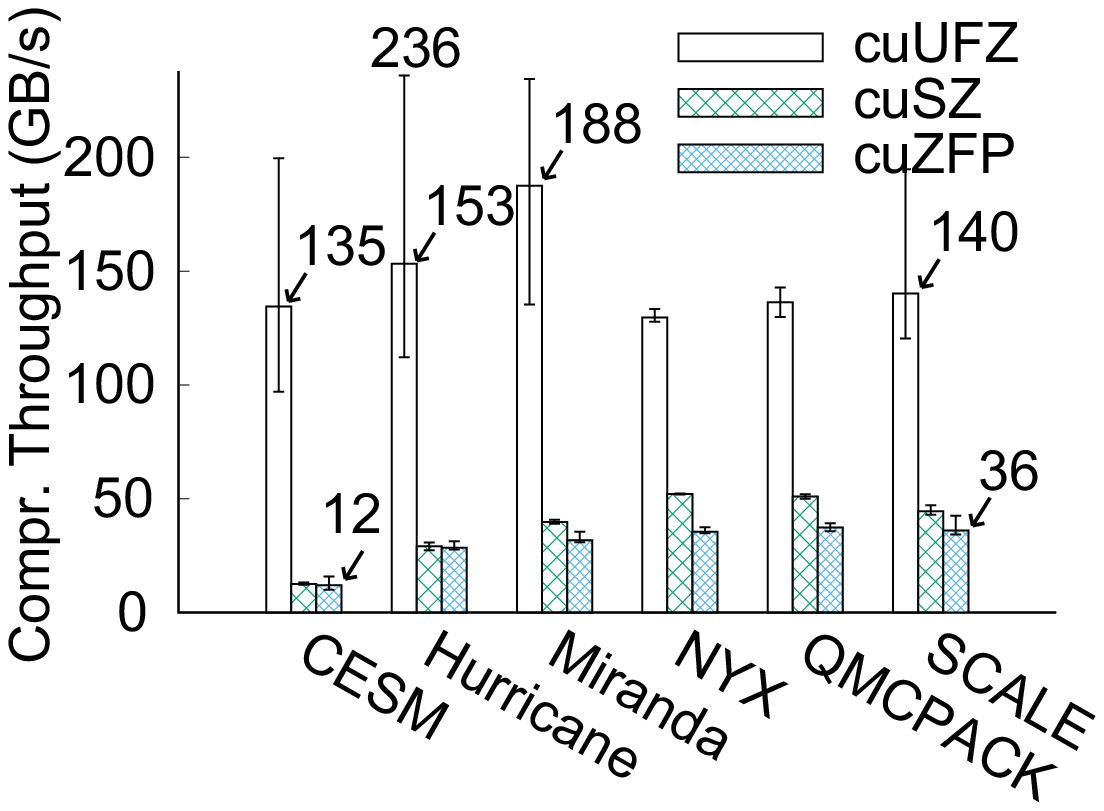}}
}
\hspace{-12mm}
\vspace{-1mm}
\caption{Overall Compression Throughput Per GPU (GB/s)}
\label{fig:gpu-CT}
\end{figure}

\begin{figure}[ht] 
\centering
\hspace{-12mm}
\subfigure[{thetaGPU (A100)}]
{
\raisebox{-1cm}{\includegraphics[scale=0.38]{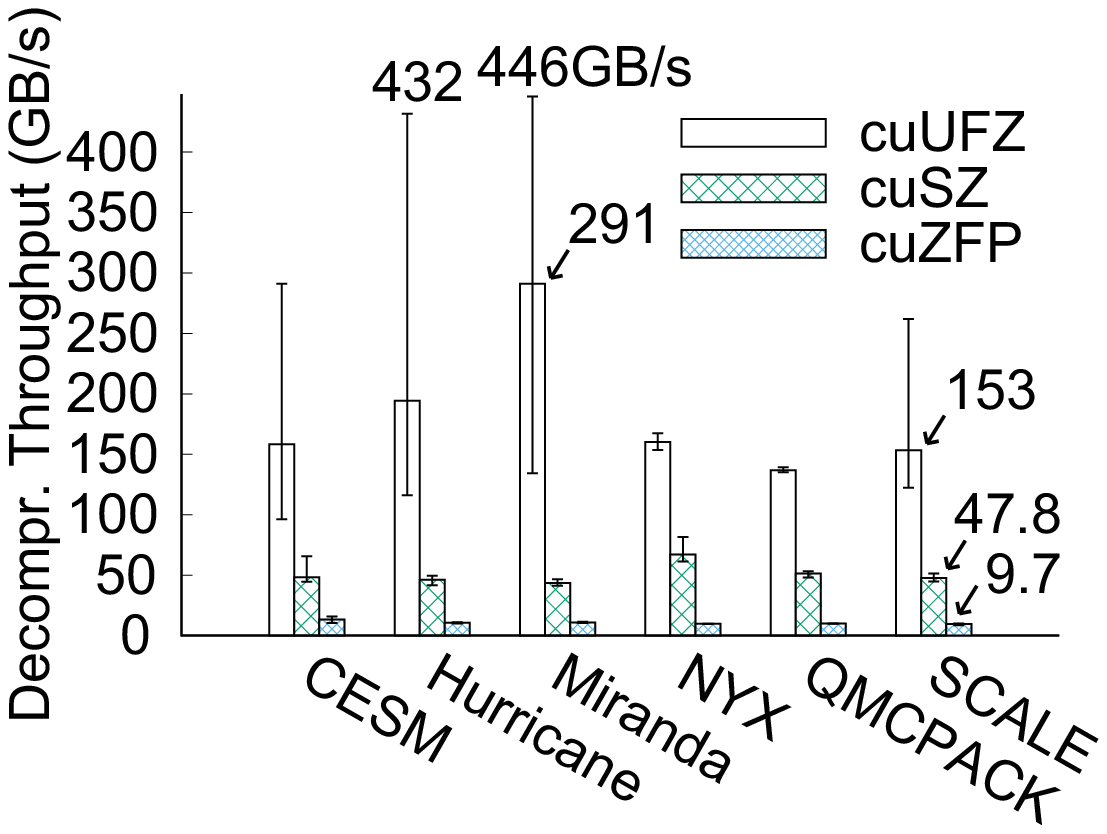}}
}
\hspace{-6mm}
\subfigure[{Summit (V100)}]
{
\raisebox{-1cm}{\includegraphics[scale=0.38]{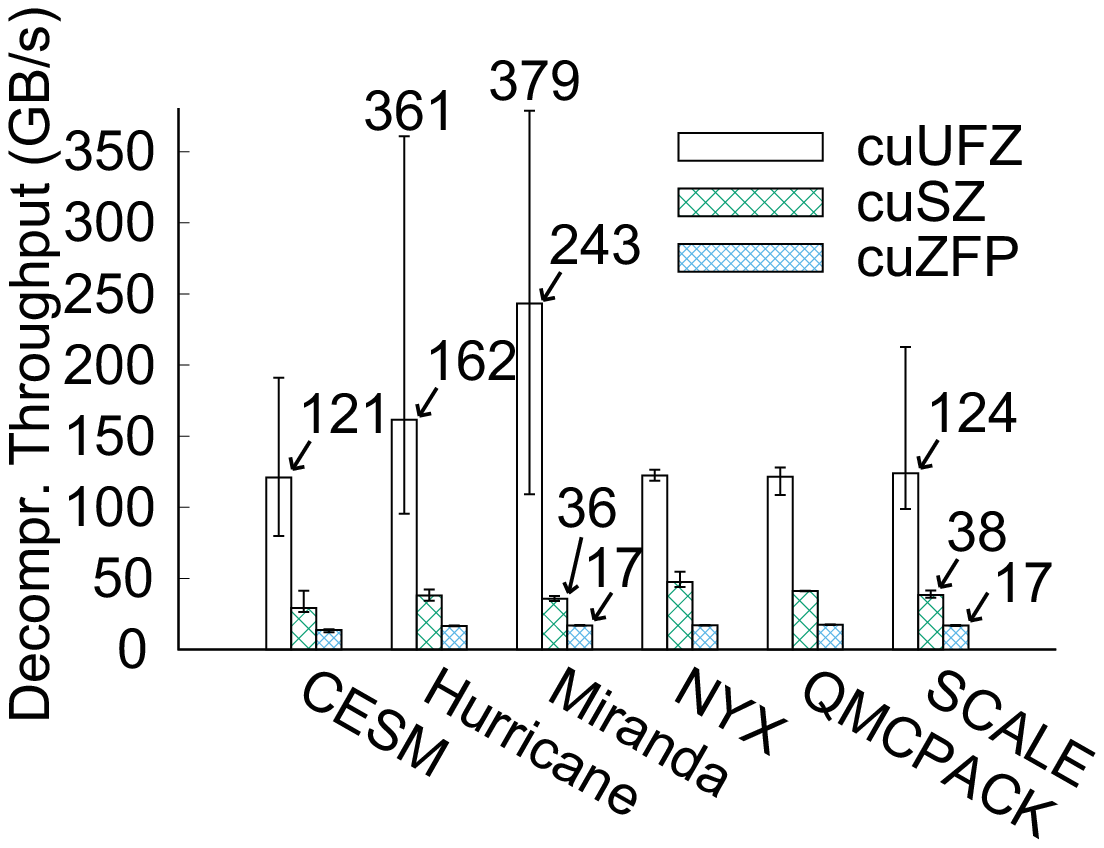}}
}
\hspace{-12mm}
\vspace{-1mm}

\caption{Overall Decompression Throughput Per GPU (GB/s)}
\label{fig:gpu-DT}
\end{figure}

We evaluate the GPU performance for cu\projectName, cuZFP, and cuSZ on two cutting-edge supercomputers -- ANL thetaGPU (A100) and ORNL Summit (V100), respectively. The compression and decompression performance results regarding all the fields of each application are presented in Figure \ref{fig:gpu-CT} and Figure \ref{fig:gpu-DT}, respectively. 

According to Figure \ref{fig:gpu-CT}, the peak performance of {\projectName} can reach up to 264GB/s (see Hurricane's result in Figure \ref{fig:gpu-CT} (a)). The overall compression performance of {\projectName} is 150$\sim$216GB/s on thetaGPU and 140$\sim$188GB/s on Summit. By comparison, both cuSZ and cuZFP suffer from very low GPU performance (9.8$\sim$86GB/s on thetaGPU and 12$\sim$52GB/s on Summit).

According to Figure \ref{fig:gpu-DT}, the peak performance of {\projectName} can reach up to 446GB/s (see Miranda's result in Figure \ref{fig:gpu-DT} (a)). The overall decompression performance of {\projectName} is 150$\sim$291GB/s on thetaGPU and 120$\sim$243GB/s on Summit. By comparison, both cuSZ and cuZFP suffer from much lower decompression performance on GPU (9.7$\sim$67GB/s on thetaGPU and 13.7$\sim$48GB/s on Summit).

\begin{figure}[ht] 
\centering
\hspace{-12mm}
\subfigure[{dumping performance (1E-2)}]
{
\raisebox{-1cm}{\includegraphics[scale=0.38]{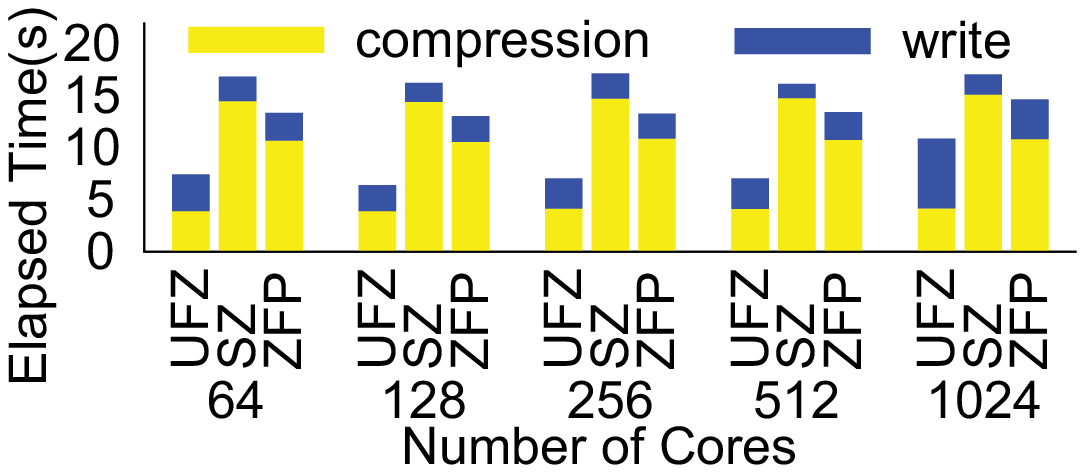}}
}
\hspace{-6mm}
\subfigure[{loading performance (1E-2)}]
{
\raisebox{-1cm}{\includegraphics[scale=0.4]{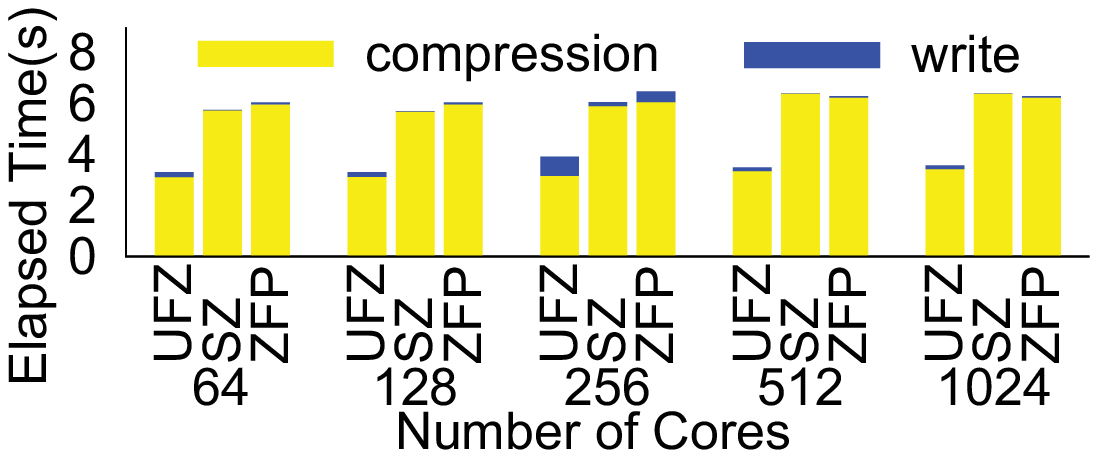}}
}
\hspace{-12mm}
\vspace{-1mm}

\hspace{-12mm}
\subfigure[{dumping performance (1E-3)}]
{
\raisebox{-1cm}{\includegraphics[scale=0.4]{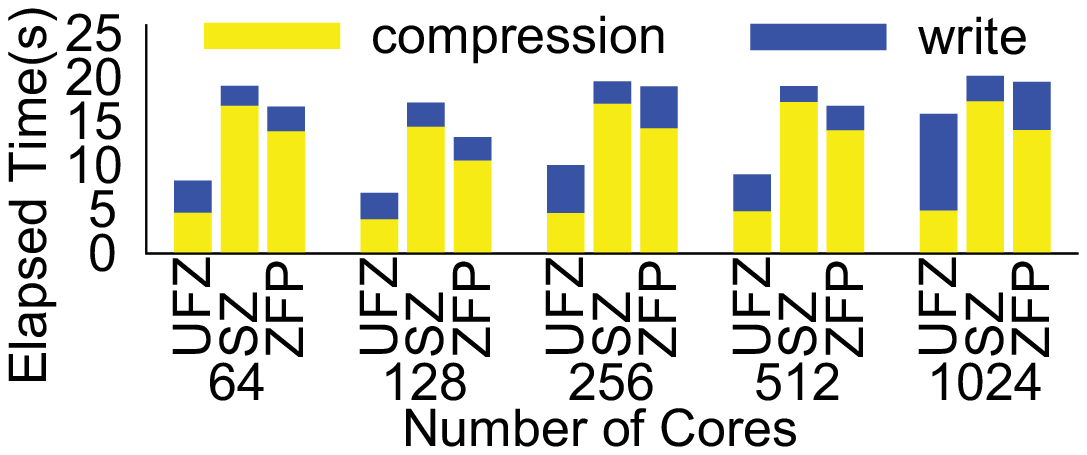}}
}
\hspace{-6mm}
\subfigure[{loading performance (1E-3)}]
{
\raisebox{-1cm}{\includegraphics[scale=0.4]{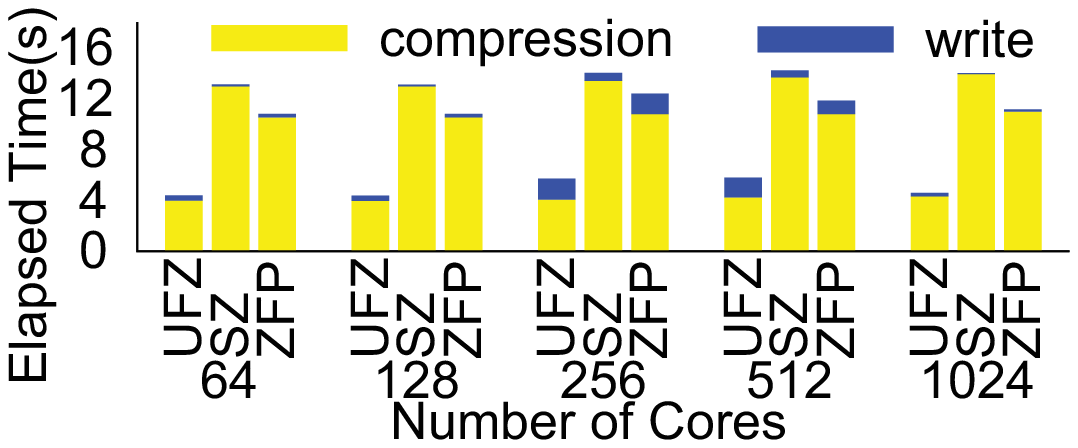}}
}
\hspace{-12mm}
\vspace{-1mm}

\hspace{-12mm}
\subfigure[{dumping performance (1E-4)}]
{
\raisebox{-1cm}{\includegraphics[scale=0.4]{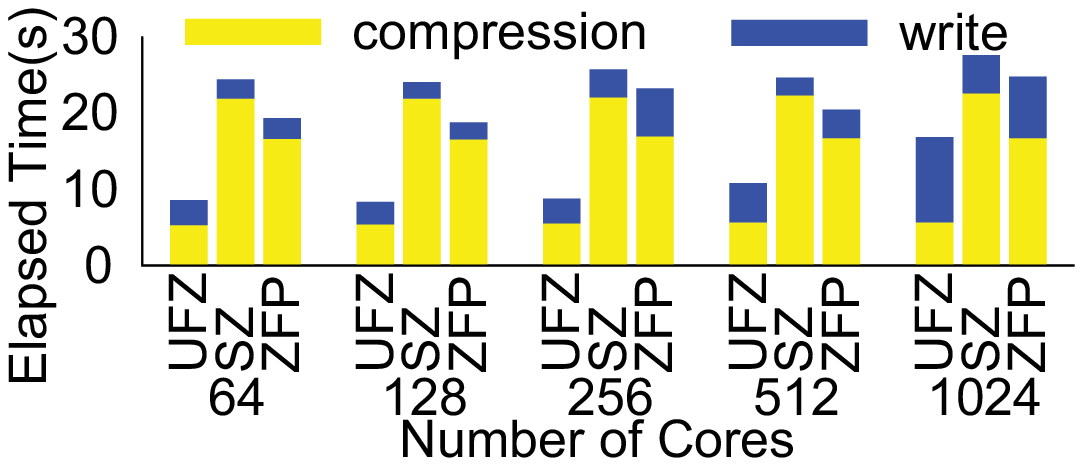}}
}
\hspace{-6mm}
\subfigure[{loading performance (1E-4)}]
{
\raisebox{-1cm}{\includegraphics[scale=0.4]{figures/load-1E-2-a.eps}}
}
\hspace{-12mm}
\vspace{-1mm}

\caption{Data Dumping/Loading Performance on thetaGPU (NYX dataset)}
\label{fig:test-io}
\end{figure}

We also evaluate the overall data dumping/loadling performance on ANL thetaGPU nodes with different execution scales. Specifically, as for the data dumping experiment, we use an MPI code to launch 64$\sim$1024 ranks/cores, each performing a lossy compression using NYX dataset and writing compressed data onto PFS. For the data loading experiment, each MPI rank reads the compressed data from parallel file system (PFS) and then performs decompression. We present the performance breakdown in Figure \ref{fig:test-io} in terms of different value-range based error bounds.  

Through the figure, we can clearly observe that the {\projectName} obtains the highest overall performance in both data dumping and data loading on thetaGPU. In absolute terms, the solution with {\projectName} takes only $\frac{1}{3}$$\sim$$\frac{1}{2}$ time to dump or load data than other solutions in most cases. That is, the I/O performance is improved by 100\%$\sim$200\% under {\projectName}. The key reason is that the thetaGPU has relatively fast I/O speed, so that the compression/decompression stage turns the key bottleneck at the execution scales in our experiments. 

\section{Conclusion and Future Work}
\label{sec:con}

In this paper, we propose an ultra-fast error-bounded lossy compression framework -- {\projectName}. %We rigorously confine the design of UFZ to use only super-lightweight calculations such as addition, subtraction and bitwise operations.
%We optimize the compression quality and performance for both CPU and GPU and perform comprehensive evaluations using two cutting-edge supercomputers' heterogeneous resources. The key insights are summarized as follows.
We perform comprehensive evaluations using 6 real-world scientific datasets and two cutting-edge supercomputers' heterogeneous resources. The key insights are summarized as follows.
\begin{itemize}
    \item With the same error bound, {\projectName} has reasonably lower compression ratios than SZ and ZFP does (0.3$\sim$3$\times$ lower than ZFP and 3$\sim$30$\times$ lower than SZ)  because it has no sophisticated data prediction/transform step and no expensive encoding algorithms such as Huffman encoding.
    \item On CPU: with the same error bound, {\projectName} is 2.5$\sim$5$\times$ as fast as ZFP and 5$\sim$7$\times$ as fast as SZ in compression; {\projectName} is 2$\sim$4$\times$ as fast as both SZ and ZFP in decompression.
    \item On GPU: with the same error bound, {\projectName}'s peak performance in compression and decompression on single GPU can reach up to 264GB/s and 446GB/s, respectively. This is 2$\sim$16$\times$ as fast as SZ and ZFP on GPU.
    \item When compressing\&writing compressed data to parallel file system (PFS) or reading\&decompressing compressed data from PFS on ANL ThetaGPU with 64$\sim$1024 cores, the overall data dumping/loading performance under {\projectName} is higher than that with SZ or ZFP by 100\%$\sim$200\%, because of {\projectName}'s ultra-fast compression and decompression and relatively fast I/O speed on ThetaGPU.
\end{itemize}
In the future work, we plan to explore how to further improve compression ratios for {\projectName}.

\section*{Acknowledgements}

This research was supported by the Exascale Computing Project (ECP), Project Number: 17-SC-20-SC, a collaborative effort of two DOE organizations – the Office of Science and the National Nuclear Security Administration, responsible for the planning and preparation of a capable exascale ecosystem, including software, applications, hardware, advanced system engineering and early testbed platforms, to support the nation’s exascale computing imperative. The material was supported by the U.S. Department of Energy, Office of Science, and by DOE’s Advanced Scientific Research Computing Office (ASCR) under contract DE-AC02-06CH11357, and supported by the National Science Foundation under Grant No. 2003709 and OAC-2104023. We acknowledge the computing resources provided on ThetaGPU, which is operated by the Argonne Leadership Computing Facility at Argonne National Laboratory.

\bibliographystyle{IEEEtran}
\bibliography{references}

\end{document}